\documentclass[11pt]{article}
\usepackage{amsmath,amsthm,amsfonts}
\usepackage[margin=1in,nohead]{geometry}
\usepackage{xcolor,url}
\usepackage{xspace,mathtools}
\usepackage{enumerate}
\usepackage{hyperref}
\usepackage{float}
\hypersetup{
    colorlinks=true, 
    linktoc=all,     
    linkcolor=blue,  
    citecolor=red,
}

\newcommand{\fullversion}[1]{#1}

\newcommand{\maybe}[1]{}

\newcommand{\ifma}[1]{#1} 

\newcommand{\coNP}{{\sf coNP}}
\newcommand{\NP}{{\sf NP}}

\newcommand{\NTime}{{\sf NTime}}
\newcommand{\NPpoly}{{\sf NP/ poly}}
\newcommand{\Ppoly}{{\sf P/ poly}}

\newcommand{\TFNP}{{\sf TFNP}}
\newcommand{\FNP}{{\sf FNP}}
\renewcommand{\P}{{\sf P}}
\newcommand{\PPAD}{{\sf PPAD}}
\newcommand{\BPP}{{\sf BPP}}

\newcommand{\PPT}{{\sf PPT}}

\newcommand{\PSPACE}{{\sf PSPACE}}
\newcommand{\IP}{{\sf IP}}

\newcommand{\PSamp}{{\sf PSamp}}
\newcommand{\HeurBPP}{{\sf HeurBPP}}
\newcommand{\HeurdBPP}{{\sf Heur}_\delta{\sf BPP}}
\newcommand{\sHeurBPP}{{\sf SearchHeurBPP}}
\newcommand{\sHeurdBPP}{{\sf SearchHeur}_\delta{\sf BPP}}
\newcommand{\remove}[1]{}

\newcommand{\Inv}{\mathsf{Inv}}

\newcommand{\view}{\mbox{\bf View}}
\newcommand{\ideall}{\textsc{IDEAL}}
\newcommand{\tideall}{\widetilde{\mathsf{IDEAL}}}

\newcommand{\Expt}{\mathbf{Expt}}
\newcommand{\expt}{\mathsf{Expt}}
\newcommand{\tExpt}{\widetilde{\mathbf{Expt}}}
\newcommand{\texpt}{\widetilde{\mathsf{Expt}}}

\newcommand{\treall}{\widetilde{\mathsf{REAL}}}

\newcommand{\bit}{\{0,1\}}

\newcommand{\ppt}{{\sf PPT}}

\newcommand{\Gen}{\mathsf{Gen}}

\newcommand{\widr}{\widetilde{r}}

\newcommand{\hatr}{\widehat{r}}

\newcommand{\widp}{\widetilde{p}}

\newtheorem{thm}{Theorem}[section]      
\newcommand{\BT}{\begin{thm}}   \newcommand{\ET}{\end{thm}}
\newtheorem{dfn}[thm]{Definition}      %
\newcommand{\BD}{\begin{dfn}}   \newcommand{\ED}{\end{dfn}}
\newtheorem{corr}[thm]{Corollary}      %
\newcommand{\BCR}{\begin{corr}} \newcommand{\ECR}{\end{corr}}
\newtheorem{Ithm}{Theorem}[section]     
\newcommand{\BIT}{\begin{Ithm}}   \newcommand{\EIT}{\end{Ithm}}
\newtheorem{lem}{Lemma}[section]  
\newcommand{\BL}{\begin{lem}}   \newcommand{\EL}{\end{lem}}
\newtheorem{prop}[lem]{Proposition}
\newcommand{\BP}{\begin{prop}}   \newcommand{\EP}{\end{prop}}
\newtheorem{clm}{Claim}         %
\newcommand{\BCM}{\begin{clm}}   \newcommand{\ECM}{\end{clm}}
\newtheorem{sclm}{SubClaim}[clm]           %
\newcommand{\BSCM}{\begin{sclm}}   \newcommand{\ESCM}{\end{sclm}}

\newtheorem{fact}[lem]{Fact}            %
\newcommand{\BF}{\begin{fact}}   \newcommand{\EF}{\end{fact}}
\renewenvironment{proof}{\noindent{\bf Proof:~~}}{\qed}
\newcommand{\BPF}{\begin{proof}} \newcommand {\EPF}{\end{proof}}
\newtheorem{prot}{Protocol}      
\newcommand{\BPR}{\begin{prot}}   \newcommand{\EPR}{\end{prot}}
\newenvironment{cproof}{\noindent{\bf Proof:~~}}{\hfill $\Box$}
\newcommand{\BCPF}{\begin{cproof}} \newcommand {\ECPF}{\end{cproof}}

\newtheorem{remark}[lem]{Remark}            %
\newcommand{\BR}{\begin{remark}}   \newcommand{\ER}{\end{remark}}

\newcommand{\BFT}{\begin{fact}}   \newcommand{\EFT}{\end{fact}}

\newcommand{\BDE}{\begin{description}}
\newcommand{\EDE}{\end{description}}
\newcommand{\BE}{\begin{enumerate}}
\newcommand{\EE}{\end{enumerate}}
\newcommand{\BI}{\begin{itemize}}
\newcommand{\EI}{\end{itemize}}
\newcommand{\BEQ}{\begin{eqnarray*}}
\newcommand{\EEQ}{\end{eqnarray*}}
\def\blackslug
{\hbox{\hskip 1pt\vrule width 8pt height 8pt depth 1.5pt\hskip 1pt}}
\def\qed{\quad\blackslug\lower 8.5pt\null\par}

\newcommand{\cD}{{\cal D}}

\newcommand{\cH}{{\cal H}}

\newcommand{\cR}{{\cal R}}

\newcommand{\cU}{{\cal U}}

\newcommand{\NN}{\mathbb{N}}

\newcommand{\poly}{\mathsf{poly}}

\newenvironment{boxfig}[2]{\begin{figure}[#1]\fbox{\begin{minipage}{\columnwidth}
                        \vspace{0.2em}
                        \makebox[0.025\columnwidth]{}
                        \begin{minipage}{0.95\columnwidth}
            {\small{
                        #2 }}
                        \end{minipage}
                        \vspace{0.2em}
                        \end{minipage}}}{\end{figure}}

\newcommand{\la}{\leftarrow}
\newcommand{\inter}[1]{\langle #1 \rangle}
\newcommand{\Chal}{\mathcal{C}}
\newcommand{\Adv}{\mathcal{A}}
\newcommand{\np}{\widetilde{\Adv}}

\newcommand{\nv}{\widetilde{\Chal}}
\newcommand{\eps}{\epsilon}

\newcommand{\bitset}{\bit}
\newcommand{\Ideal}{\mathsf{PInv}}
\newcommand{\PInv}{\mathsf{PInv}}

\newcommand{\Succ}{\mathsf{Succ}}
\newcommand{\KL}{\textbf{KL}}
\newcommand{\Exp}{\mbox{E}}

\newcommand{\defsuc}[2]{\Pr[\inter{#1,\Chal}(1^{#2})=1]}

\newcommand{\ttp}{\widetilde{P}}
\newcommand{\tv}{\widetilde{V}}
\newcommand{\myprot}[2]{#2}
\newcommand{\promise}{promise-true}
\newcommand{\cPromise}{Promise-true}

\begin{document}

\begin{titlepage}
  
\title{\emph{Is it Easier to Prove Theorems that are Guaranteed to be True?}}

\author{Rafael Pass\thanks{Cornell Tech. Supported in part by NSF Award SATC-1704788, NSF Award RI-1703846, and AFOSR Award FA9550-18-1-0267.
This research is based upon work supported in part by the Office of
the Director of National Intelligence (ODNI), Intelligence Advanced
Research Projects Activity (IARPA), via 2019-19-020700006.  The views
and conclusions contained herein are those of the authors and should
not be interpreted as necessarily representing the official policies,
either expressed or implied, of ODNI, IARPA, or the U.S. Government.
The U.S. Government is authorized to reproduce and distribute reprints
for governmental purposes notwithstanding any copyright annotation
therein.} \\Cornell Tech \\ \texttt{rafael@cs.cornell.edu}
\and Muthuramakrishnan Venkitasubramaniam\thanks{Supported by Google Faculty Research Grant, NSF Award CNS-1618884 and Intelligence Advanced Research Projects Activity (IARPA) via 2019-19-020700009. Work done partially at Cornell Tech sponsored by Cornell Tech and DIMACS Research Visit Program via DIMACS/Simons Collaboration in Cryptography.} \\ University of Rochester \\ \texttt{muthuv@cs.rochester.edu}}
 \maketitle
  
\begin{abstract}
\noindent Consider the following two fundamental open problems in complexity theory:
\BI
\item Does a  hard-on-average language in $\NP$
  imply the existence of one-way functions?
  \item Does a  hard-on-average language in $\NP$
    imply a hard problem in $\TFNP$ (i.e., the class of \emph{total} $\NP$ search
    problem)?
  \EI
  Our main result is that the answer to (at least) one of these questions is yes.
%
%

  Both one-way functions and problems in $\TFNP$ can
be interpreted as \emph{\promise}
distributional $\NP$ search problems---namely, distributional search
problems where the sampler only samples \emph{true} statements.
As a direct corollary of the above
result, we thus get that the existence of a hard-on-average distributional $\NP$ search problem
implies a hard-on-average {\promise} distributional $\NP$
search problem. In other words,
\begin{quote}
  \emph{It is no easier to find witnesses (a.k.a. proofs) for
    efficiently-sampled statements (theorems) that are guaranteed to be true.}
  \end{quote}

This result follows from a more general study of \emph{interactive
puzzles}---a generalization of average-case hardness
in $\NP$---and in particular, a novel round-collapse
theorem for computationally-sound protocols, analogous to Babai-Moran's
celebrated round-collapse theorem for information-theoretically sound protocols.
As another consequence of this treatment, we show that the existence of $O(1)$-round
public-coin \emph{non-trivial}
arguments (i.e., argument systems that are not proofs) imply the existence
of a hard-on-average problem in $\NPpoly$.

\end{abstract}
\end{titlepage}
\fullversion{

\tableofcontents 
\newpage}

\section{Introduction}
\setcounter{page}{1}
Even
if $\NP \neq \P$, it could be that \emph{in practice}, $\NP$ problems
are easy in the sense that the problems we encounter in ``real life'' come
from some distribution that make them easy to solve.
The complexity-theoretic study of average-case hardness of $\NP$ problems
addresses this problem \cite{L86,G91,Ben-DavidCGL92,ImpagliazzoL90}. A particularly 
appealing abstraction of an average-case analog of $\NP \neq
\P$ was provided by Gurevich in his 1989 essay \cite{G89} through his
notion of a \emph{Challenger-Solver Game}.\footnote{Gurevich actually
  outlines several classes of Challenger-Solver games; we here outline
  one particular instance of it, focusing on $\NP$ search problems.}  Consider
a probabilistic polynomial-time \emph{Challenger} $C$ who samples an instance
$x$ and provides it to the \emph{Solver} $S$. The solver $S$ is
supposed to find a witness to $x$ and is said to win if either (1) the
statement $x$ chosen by the challenger is false, or (2) $S$ succeeds in
finding a witness $w$ for $x$. We refer to the Challenger-Solver
game as being \emph{hard} if no probabilistic polynomial-time (PPT) solver succeeds in
winning in the game with inverse polynomial probability. (In
  other words, such a game models a hard-on-average distributional search problem in $\NP$.)
The existence of a hard Challenger-Solver game means that there exists
a way to efficiently sample mathematical statements $x$ that no computationally
bounded mathematician can find proofs for. (Impagliazzo \cite{Imp95}
considers a similar type of game between Professor Grauss and young
Gauss, where Professor Grauss is trying to embarrass Gauss by picking
mathematical problems that Gauss cannot solve.)

But, an unappealing aspect of a Challenger-Solver game (which already goes
back to the definition of distributional search problems \cite{Ben-DavidCGL92}) is that
checking whether the solver wins cannot necessarily be efficiently
done, as it requires determining whether the sampled instance $x$ is
in the language. Does it make the problem easier if we restrict the
challenger to \emph{always} sample true statements
$x$?\footnote{Or equivalently, to distributions where one can
  efficiently check when the sampler outputs a false instance.} In other
words,
  \emph{``Is it easier to find proofs for efficiently-sampled mathematical statements that
    are guaranteed to be true?''}
  In complexity-theoretic terms: 
\begin{quote}
\emph{Does the existence of an hard-on-average distributional search
  problem in
$\NP$ imply the existence of a hard-on-average distributional search
problem where the sampler {\bf only samples true statements}?}
\end{quote}
We refer to distributional search
problems where the sampler only samples true statements as
\emph{\promise} distributional search
problems.
The above question, and the notion of a {\promise}
distributional search problems, actually predates the formal study of
average-case complexity:
It was noted already by Even, Selman and Yacobi
\cite{ESY} in 1984 that for typical applications of (average-case) hardness for $\NP$ problems---in particular,
for cryptographic applications---we need hardness for instances that
are ``promised'' to be true. As they noted (following
\cite{EY}\footnote{As remarked in \cite{EY}, these type of ``problems
  with a promise'' can be traced back even further: they are closely related to what was referred to as
  a ``birdy'' problem in \cite{ginsburg} and a ``partial algorithm problem'' in
  \cite{Ullian67}, in the study of context-free languages}), in the context of public-key encryption, security
is only required for ciphertexts that are sampled as valid encryptions of
some message. (This motivated \cite{ESY} to introduce the concept of a
promise problem; see also \cite{Gpromise} for further discussion on this
issue and the connection to average-case complexity.)

Intuitively, restricting to challengers that only sample true
statements ought to make the job of the challenger a lot harder---it
now needs to be sure that the sampled instance is true. There
are two natural methods for the challenger to achieve this task:
\begin{itemize}
  \item [(a)] sampling the statement $x$ together with a witness $w$ (as
    this clearly enables the challenger to be sure that $x$ is true);
    and,
    \item [(b)] restricting to $\NP$ languages where \emph{every} statement is
      true.
    \end{itemize}
As noted
by Impagliazzo \cite{G89,Imp95}, the existence of a challenger-solver game
satisfying restriction (a) is equivalent to the existence of one-way
functions.\footnote{That is, a function $f$ that can be computed in
  polynomial time but cannot be efficiently inverted. Such a function $f$
  directly yields the desired sampling method: pick a random string
  $r$ and let $x = f(r)$ be the statement and $r$ the
  witness. Conversely, to see why the existence of such a sampling
  method implies a one-way function, consider the
function $f$ that takes the random coins used by the sampling method and
outputs the instance generated by it.}
But whether the existence of a hard-on-average
language in $\NP$ implies the existence of one-way functions is arguably
the most important open problem in the foundations of Cryptography: One-way functions are both necessary
\cite{ImpagliazzoL89} and sufficient for
many of the central cryptographic tasks (e.g., pseudorandom generators
\cite{HILL99}, pseudorandom functions \cite{GGM84}, private-key encryption
\cite{GM84, BM88}). As far as we know, there are only two approaches towards demonstrating
the existence of one-way functions from average-case $\NP$ hardness: (1) Ostrovsky and
Wigderson \cite{OstrovskyW93} demonstrate such an implication assuming
that $\NP$ has zero-knowledge proofs \cite{GMW91}, (2) Komargodski et
al. \cite{KomargodskiMNPRY14} demonstrate the implication (in fact, an even stronger
implication, showing worst-case hardness of $\NP$ implies one-way
functions) assuming the existence of \emph{indistinguishability
  obfuscators} \cite{BarakGIRSVY01}. Both of these additional assumptions are
not known to imply one-way functions on their own (in fact, they
are unconditionally true if $\NP \subseteq \BPP$).

A hard challenger-solver game
satisfying restriction (b), on the other hand, is syntactically equivalent to a
hard-on-average problem in the class $\TFNP$ \cite{MegiddoP91}: the class $\TFNP$ (total function $\NP$)
is the search analog of $\NP$ with the additional
guarantee that \emph{any} instance has a solution. In other words,
$\TFNP$ is the 
class of search problems in $\NP \cap \coNP$ (i.e., $F(\NP \cap \coNP)$).
In recent years, $\TFNP$ has
attracted extensive attention due to its natural syntactic subclasses
that capture the computational complexity of important search problems
from algorithmic game theory, combinatorial optimization and
computational topology---perhaps most notable among those are the classes
$\PPAD$ \cite{Papadimitriou94,GP16}, which characterizes the hardness of
computing Nash equilibrium
\cite{DaskalakisGP09,ChenDT09,DaskalakisP11},
and $\mathsf{PLS}$ \cite{JohnsonPY85}, which characterizes the hardness
of local search.
A central open problem is whether
(average-case) $\NP$ hardness implies (average-case) $\TFNP$ hardness.
A recent elegant result by Hubacek, Naor, and Yogev \cite{HubacekNY17} shows
that under certain strong ``derandomization''
assumptions \cite{NW94,IW97,MiltersenV05,BarakOV07}---the existence of 
Nisan-Wigderson (NW) \cite{NW94} type pseudorandom generators that fool circuits with oracle gates to languages in the second
level of the polynomial hierarchy\footnote{Such PRGs are known under
  the assumption that $E=DTIME[2^{O(n)}]$ has no $2^{\epsilon n}$
  sized $\Pi_2$-circuits, for all $\epsilon > 0$, where a
  $\Pi_2$-circuit is a standard circuit that can additionally perform
  oracle queries to any language $L \in \Pi_2$ (i.e., any language in
  the second level of the polynomial hierarchy).}---(almost everywhere) average-case hardness of $\NP$ implies
average-case hardness of $\TFNP$.\footnote{\cite{HubacekNY17} also show that
  average-case hardness of $\NP$ implies an average-case hard problem
  in $\TFNP/poly$ (i.e,. $\TFNP$ with a \emph{non-uniform
    verifier}). In essence, this follows since non-uniformity enables 
  unconditional derandomization.}
Hubacek et al. also present another condition under which $\TFNP$ is
average-case hard: assuming the existence of one-way functions and 
\emph{non-interactive witness indistinguishable proofs (NIWI)}
\cite{FS90,DworkN00,BarakOV07} for $\NP$.

The above mentioned works thus give complexity-theoretic assumptions
(e.g., the existence of zero-knowledge proofs for $\NP$, or strong
derandomization assumption) under which the above problem has a
positive resolution. But these assumptions are both complex and strong.

Our main result provides a resolution to the above problem
\emph{without any complexity-theoretic assumption}:\footnote{Pedantically, it is not a fully complete
  resolution as we start with an \emph{almost-everywhere} hard problem
  and only get an \emph{infinitely-often} hard problem. But, except for this
  minor issue, it is a complete resolution. We also note that earlier
  results \cite{OstrovskyW93,HubacekNY17} also require starting off
  with an almost-everywhere hard-on-average language in $\NP$.}
\BT [Informally stated] \label{thm1} The existence of an
  almost-everywhere hard-on-average language in $\NP$\footnote{That is, a language
  in $\NP$ such that for every $\delta>0$, no PPT attacker $A$ can decide random instances with
  probability greater than $\frac{1}{2} + \delta$ for
    \emph{infinitely many} (as opposed to all) $n \in N$. Such an
  ``almost-everywhere'' notion is more commonly used in the
  cryptographic literature.} implies the
  existence of a hard-on-average {\promise} distributional search problem in
  $\NP$.
  \ET

  In fact, we demonstrate an even stronger statement. Perhaps
  surprisingly, we show that
  without loss of generality, the sampler/challenger of the
  distributional search problem needs to satisfy one of the
  above two ``natural'' restrictions:
\BT [Informally stated] \label{thm2} The existence of an
  almost-everywhere hard-on-average language in $\NP$ implies either
  (a) the existence one-way functions, or (b) a hard-on-average
  $\TFNP$ problem.
  \ET
  In other words, in Impagliazzo's Pessiland \cite{Imp95} (a world where $\NP$ is
hard-on-average, but one-way functions do not exist), $\TFNP$ is
unconditionally hard (on average).

Towards proving this result, we consider an alternative notion of a
Challenger-Solver game, which we refer to as a \emph{Interactive Puzzle}. Roughly speaking, there are 2 
differences: (1) whether the solver wins should always be
computationally feasible to determine, and (2) we allow for more than
just 2 rounds of interaction. As we hope to convey, the study of
interactive puzzles is intriguing in its own right and yields other applications.

    \subsection{Interactive Puzzles}
We initiate a complexity-theoretic study of \emph{interactive 
puzzles}: 2-player interactive games between a
polynomial-time challenger $\Chal$
and an Solver/Attacker\footnote{Following the nomenclature in the
  cryptographic literature, we use the name Attacker instead of Solver.}
  satisfying the following properties:
\BDE
\item [Computational Soundness:] There does not exist a
  \emph{probabilistic polynomial-time (PPT)} attacker $\Adv^*$ and
  polynomial $p$
  such that $\Adv^*(1^n)$ succeeds in making $\Chal(1^n)$ output 1 with
  probability $\frac{1}{p(n)}$ for all sufficiently large $n \in N$.
  \item [Completeness/Non-triviality:] There exists a negligible function $\mu$ and an \emph{inefficient} attacker
  $\Adv$ that on input $1^n$ succeeds in making $\Chal(1^n)$ output 1 with probability $1-\mu(n)$
  for all $n \in N$.
  \item [Public Verifiability:] Whether $\Chal$ accepts should just be
    a deterministic function of the transcript.
\EDE
In other words, (a) no \emph{polynomial-time} attacker,
$\Adv^*$, can make $C$ output 1 with inverse polynomial
probability, yet (b) there exists a \emph{computationally  unbounded}
attacker ${\Adv}$ that makes
$C$ output 1 with overwhelming probability. 
We refer to $\Chal$ as a
\emph{$k(\cdot)$-round computational puzzle} (or simply a
$k(\cdot)$-round puzzle) if $\Chal$ satisfies the above completeness
and computational soundness conditions, while restricting $\Chal(1^n)$ to communicate with
$\Adv$ in $k(n)$ rounds. In this work, we mostly restrict our attention
to \emph{public-coin} puzzles, where the Challenger's messages are
simply random strings.

As an example of a 2-round public-coin puzzle, let $f$ be a one-way permutation
and consider a game where $\Chal(1^n)$ samples a random $y \in \{0,1\}^n$ and
  requires the adversary to output a preimage $x$ such that $f(x) =
  y$. Since $f$ is a permutation, this puzzle has ``perfect''
  completeness---an unbounded attacker ${\Adv}$ can always find a
  pre-image $x$. By the one-wayness of $f$ (and the permutation
  property of $f$), we also have that no $\PPT$ adversary $\Adv^*$ can
  find such an $x$ (with inverse polynomial probability), and thus soundness holds.
If however, $f$ had only been a one-way function and not a
permutation, then we can no longer sample a uniform $y$, but rather
must have $\Chal$ first sample a random $x$ and next output $y =
f(x)$. This 2-round puzzle does not satisfy the public-coin property,
but it still have perfect completeness.

  Its not hard to see that the existence of 2-round (public-coin) puzzles is ``essentially'' equivalent to the existence of an average-case
hard problem in $\NP$: any 2-round public-coin puzzle trivially implies a hard-on-average search problem (w.r.t. the uniform
distribution) in
$\NP$ and thus by \cite{ImpagliazzoL90} also a hard-on-average decision problem in
$\NP$. Furthermore, ``almost-everywhere''
hard-on-average languages in $\NP$
also imply
the existence of a 2-round puzzle (by simply sampling many
random instances $x$ and asking the attacker to provide a witness for at least, say, $1/3$ of the instances).%
\footnote{The reason we need the language to be
  \emph{almost-everywhere} hard-on-average is to guarantee that YES
  instances exists for every sufficiently large input length, or else
  completeness would not hold.}

\BP [informally stated] The existence of an (almost-everywhere) hard-on-average language
in $\NP$ implies the existence of a 2-round puzzle. Furthermore, the
existence of a 2-round puzzle implies the existence of a
hard-on-average language in $\NP$.
\EP
Thus, 2-round puzzles are ``morally''
(up to the infinitely-often/almost-everywhere issue)
equivalent to the existence of a hard-on-average language in $\NP$.
As such, $k(\cdot)$-round puzzles are a natural way to generalize
average-case hardness in $\NP$. Additionally, natural restrictions of 
2-round puzzles capture natural subclasses of distributional problems in $\NP$:
\BI
\item the existence of a
hard-on-average problem in $\TFNP$ is syntactically equivalent to the
existence of a 2-round \emph{public-coin} puzzle \emph{with perfect completeness}.
\item the existence of a hard-on-average \emph{\promise}
  distributional search problem is syntactically equivalent to a
  2-round (private-coin) puzzle \emph{with perfect completeness}.
  \EI
  
While the game-based modeling in the notion of a puzzle is common in the cryptographic
literature---most notably, it is commonly used to model cryptographic
assumptions \cite{Naor03,Pass11,GentryW11}, complexity-theoretic
consequences or properties of puzzles have remained largely
unexplored.

\subsection{The Round-Complexity of Puzzles}
Perhaps the most basic question regarding the existence of interactive
puzzles is whether the existence of a $k$-round puzzle is actually a weaker
assumption than the existence of a $k-1$ round puzzle. In particular,
do interactive puzzles actually generalize beyond
just average-case hardness in $\NP$:
\begin{quote}
\emph{Does the existence of a $k$-round puzzle imply the
  existence of $(k-1)$-round puzzle?}
\end{quote}
We here focus our attention only on public-coin puzzles.
At first sight, one would hope the classic ``round-reduction'' theorem
due to Babai-Moran (BM) \cite{BM88} can be applied to collapse any
$O(1)$-round puzzle into a 2-round puzzle (i.e., a hard-on-average $\NP$ problem). Unfortunately, while BM's round
reduction technique indeed works for all \emph{information-theoretically}
sound protocols, Wee \cite{Wee06} demonstrated that BM's round reduction
fails for computationally sound protocols. In particular, Wee shows
that black-box proofs of security cannot be used to prove that BM's
transformation preserves soundness even when applied to just 3-round
protocols, and demonstrates (under computational assumptions) a concrete 4-round protocol for which BM's
round-reduction results in an unsound protocol.

As BM's round reduction is the only known round-reduction technique
(which does not rely on any assumptions), it was generally conjectured
that the existence of a $k$-round puzzle is a strictly stronger assumption
than the existence of a $(k+1)$-round puzzle---in particular, this
would imply the
existence of infinitely many worlds between Impagliazzo's Pessiland
and Heuristica \cite{Imp95}
(i.e., infinitely many worlds where $\NP \neq P$ yet average-case $\NP$ hardness does not exist).
Further evidence in this direction comes from a work by Gertner et al. \cite{GertnerKMRV00} which shows a black-box separation between
$k$-round puzzles and $(k+1)$-round puzzles for a
particular cryptographic task (namely that of a key-agreement
scheme).\footnote{The example from \cite{GertnerKMRV00} isn't quite
  captured by our notion of a computational puzzle as their challenger is not
  public coin.}

In contrast to the above negative results, our main technical result provides an
affirmative answer to the above question---we demonstrates a round-reduction
theorem for puzzles.
\BT [informally stated] \label{thm:babaimoran} 
For every constant $c$, the existence of a $k(\cdot)$-round
public-coin puzzle
is equivalent to the existence of a $(k(\cdot)-c)$-round public-coin puzzle.
\ET
\noindent In particular, as corollary of this result, we get 
that the assumption that a $O(1)$-round public-coin puzzle exists is \emph{not} weaker than
the assumption that average-case
hardness in $\NP$ exists:
\BCR [informally stated]
The existence of an $O(1)$-round puzzle implies 
the existence of a hard-on-average problem in $\NP$.
\ECR
\noindent Perhaps paradoxically, we strongly rely on BM's round reduction
technique, yet we rely on a \emph{non-black-box} security
analysis. Our main technical lemma shows that if \emph{infinitely-often
  one-way functions}\footnote{Recall that a \emph{one-way
      function} $f$ is a function that is efficiently computable, yet there
  does not exist a PPT attacker $A$ and polynomial
  $p(\cdot)$ such that $A$ inverts $f$ with probability
  $\frac{1}{p(n)}$ for \emph{infinitely many} inputs lengths $n \in
  N$. A function $f$ is \emph{infinitely often one-way} if the same
  conditions hold except that we only require that no PPT attacker $A$
  succeeds in inverting $f$ with probability $\frac{1}{p(n)}$
  \emph{for all} sufficiently large
  $n\in N$---i.e., it is hard for invert $f$ ``infinitely often''} do not exist (i.e., if we can invert any
function for all sufficiently large input lengths), then BM's round
reduction actually works:
\BL [informally stated] \label{main:lem} Either infinitely-often one-way functions
exist, or BM's round-reduction transformation turns a
$k(\cdot)$-round 
puzzle into a $(k(\cdot)-1)$-round puzzle.
\EL
\noindent We provide a proof outline of Lemma \ref{main:lem} in Section
\ref{sec:bmoverview}.
The proof of Theorem~\ref{thm:babaimoran} now easily follows by considering two cases:
\begin{description}
\item {\bf Case 1: (Infinitely-often) one-way functions exists.} In
  such a world, we can rely on Rompel's construction of a universal
  one-way hashfunction \cite{NY89,Rompel90} to get a 2-round puzzle. 
\item {\bf Case 2: (Infinitely-often) one-way functions does not exist.} In
  such a world, by Lemma \ref{main:lem}, BM's round reduction preserves
  soundness of the underlying protocol and thus we have gotten a
  puzzle with one round less. We can next iterate BM's round reduction any
  constant number of times.
\end{description}

A natural question is whether we can collapse more than a constant
number of rounds. Our next result---which characterizes
the existence of $\poly(n)$-round puzzles---shows that this is
unlikely.
\BT[informally stated]  \label{thm:introexist}
For every $\epsilon>0$, there exists an $n^\epsilon$-round
(public-coin) puzzle if and only if $\PSPACE
\not\subseteq \BPP$.
\ET
\noindent In particular, if $n^\epsilon$-round public-coin puzzles imply $O(1)$-round public-coin puzzles,
then by combining Theorem \ref{thm:babaimoran} and Theorem \ref{thm:introexist}, we have that
$\PSPACE \not \subseteq \BPP$ implies the existence of a hard-on-average problem
in $\NP$, which seems unlikely. Theorem \ref{thm:introexist} also shows
that the notion of an interactive puzzle (with a super
constant-number of rounds) indeed is a non-trivial generalization of
average-case hardness in $\NP$. Theorem \ref{thm:introexist} follows
using mostly standard techniques.\footnote{Any
puzzle $\Chal$ can be
broken using a $\PSPACE$ oracle (as the optimal strategy can be found using a $\PSPACE$ oracle), so if $\PSPACE \subseteq \BPP$, it
can also be broken by a probabilistic polynomial-time algorithm. For the other direction, recall that
worst-case to average-case reductions are known for $\PSPACE$
\cite{FF93,BabaiFNW93}. In other words, there exists a language $L \in
\PSPACE$ that is hard-on-average assuming $\PSPACE \not \subseteq
\BPP$. Additionally, recall that $\PSPACE$ is closed under complement. We then construct a public-coin puzzle where $\Chal$ first
samples a hard instance for $L$ and then asks $\Adv$ to determine whether
$x \in L$ and next provide an
interactive proof---using \cite{Shamir92,LundFKN92} which is public-coin---for containment
or non containment in $L$. This puzzle clearly satisfies the
completeness condition. Computational soundness, on the other hand, follows directly
from the hard-on-average property of $L$ (and the unconditional
soundness of the interactive proof of \cite{Shamir92}).}

We next present some complexity-theoretic consequences of our
treatment of interactive puzzles.

\subsection{Achieving Perfect Completeness: Proving  Theorem \ref{thm2}}
We outline how the round-reduction theorem can be used to prove
Theorem~\ref{thm2} in the following steps: 
\BI
\item 
As mentioned
above, an (almost-everywhere) hard-on-average problem in $\NP$
yields a 2-round puzzle; 
\item We can next use a standard technique from the literature on
  interactive proofs (namely the result of \cite{FurerGMSZ89}) to turn this puzzle into a 3-round puzzle with
  \emph{perfect completeness}.
\item We next observe that the BM transformation 
  preserves perfect completeness of the protocol. Thus, by Lemma \ref{main:lem}, 
either infinitely-often one-way functions exist, or we can get a 2-round puzzle with perfect completeness.
\item Finally, as observed above, the existence of a 2-round puzzle with
  perfect completeness is syntactically equivalent to the existence
  of a hard-on-average problem in $\TFNP$ (with respect to the uniform
  distribution on instances).
\EI
The above proof approach actually only concludes a slightly weaker
form of Theorem~\ref{thm2}---we only show that either $\TFNP$ is hard or
\emph{infinitely-often} one-way functions exist. As infinitely-often
one-way functions directly imply 2-round \emph{private-coin} puzzles
with perfect completeness, which (as observed above) are syntactically equivalent
to hard-on-average \emph{\promise} distributional search
problems, this however already suffices to prove Theorem \ref{thm1}.

We can get the
proof also of the stronger conclusion of Theorem \ref{thm2} (i.e., conclude the existence of
standard (i.e., ``almost-everywhere'') one-way functions), by noting that
an almost-everywhere hard-on-average language in $\NP$ actually implies an
2-round puzzle satisfying a ``almost-everywhere'' notion of soundness,
and for such ``almost-everywhere puzzles'', Lemma \ref{main:lem} can be strengthened to show that either
one-way functions exist, or BM's round-reduction works.\footnote{More
  precisely, the variant of Lemma \ref{main:lem} says that either
  one-way functions exist, or the existence of a  $k$-round
  almost-everywhere puzzle
  yields the existence of a $k-1$-round puzzle
(with the standard, infinitely-often, notion of soundness).}

\subsection{The Complexity of Non-trivial Public-coin Arguments}
Soon after the introduction of interactive proof by Goldwasser,
Micali and Rackoff \cite{GMR89} and Babai and Moran \cite{BM88}, Brassard,
Chaum and Crepeau \cite{BrassardCC88} introduced the notion of an interactive
\emph{argument}. Interactive arguments are defined identically to
interactive proofs, but we relax the soundness condition to only hold
with respect to non-uniform $\PPT$ algorithms (i.e., no non-uniform
$\PPT$ algorithm can produce proofs of false statements, except with
negligible probability).

Interactive arguments have proven extremely useful in the
cryptographic literature, most notably due to the feasibility
(assuming the existence of collision-resistant hashfunctions) 
of \emph{succinct} public-coin argument systems for $\NP$---namely, argument
systems with sublinear, or even polylogarithmic communication
complexity \cite{Kilian92, M00}. Under widely believed complexity
assumptions (i.e., $\NP$ not being solvable in subexponential time),
interactive \emph{proofs} cannot be succinct \cite{GH98}.

A fundamental problem regarding interactive arguments involves characterizing the complexity of \emph{non-trivial} argument
systems---namely interactive arguments that are \emph{not} interactive proofs
(in other words, the soundness condition is inherently
computational).
As far as we know, the first explicit formalization of this question
appears in a recent work by Goldreich \cite{oded18}, but the notion of a non-trivial argument has been discussed in the community
for at least 15 years.\footnote{Wee \cite{Wee05} also considers a
  notion of a non-trivial argument, but his notion refers to what
  today is called a succinct argument.}

We focus our attention on \emph{public-coin} arguments (similar to our
treatment of puzzles). Using our
interactive-average-case hardness treatment, we are able to establish
an ``almost-tight'' characterization of constant-round public-coin non-trivial arguments.
\BT [informally stated]  \label{thmNT:constant}
The existence of a $O(1)$-round public-coin non-trivial argument for
any language $L$ implies a hard-on-average language in $\NPpoly$. Conversely, the
existence of a hard-on-average language in $\NP$ implies an
(efficient-prover) 2-round public-coin non-trivial argument for $\NP$.
\ET
The first part of the theorem is shown by observing that any
public-coin non-trivial argument can be turned into a
\emph{non-uniform} public-coin puzzle (where the challenger is a non-uniform
$\PPT$ algorithm), and next observing that our round-collapse theorem
also applies to non-uniform puzzles. The second part follows from the
observation that we can take any $\NP$ proof for some language $L$
and extending it into a 2-round non-trivial argument for $L$ where the verifier samples a random statement $x'$ from a
hard-on-average language $L'$ and next requiring the prover to provide
a witness $w$ that either $x \in L$ or $x' \in L'$. Completeness
follows trivially (as we can always provide a normal $\NP$ witness
proving that $x \in L$, and computational soundness follows directly if
$L'$ is sufficiently hard-on-average (in the sense that it is hard to
find witnesses to true statements with inverse polynomial probability).
This argument system is not a proof, though, since by the
hard-on-average property of $L'$, there must exist infinitely many
input lengths for which random instances are contained in $L'$ with inverse
polynomial probability.

We finally observe that the existence of $n^{\epsilon}$-round
non-trivial public-coin arguments is equivalent to $\PSPACE \not
\subseteq \Ppoly$. 

\BT[informally stated]  \label{thmi:NTexist}
For every $\epsilon>0$, there exists an (efficient-prover) $n^\epsilon$-round non-trivial
public-coin argument (for $\NP$) if and only if $\PSPACE
\not\subseteq \Ppoly$.
\ET
The ``only-if'' direction was already proven by Goldreich
\cite{oded18} and follows just as the only-if direction
of Theorem \ref{thm:introexist}. The ``if'' direction follows by combining a
standard $\NP$ proof with the puzzle from Theorem \ref{thm:introexist} (which
becomes sound w.r.t. nu $\PPT$ attacker assuming $\PSPACE
\not\subseteq \Ppoly$), and requiring the prover to either provide the
$\NP$ witness, or to provide a solution to the puzzle. 

\subsection{Proof Overview for Lemma \ref{main:lem}}\label{sec:bmoverview}
We here provide a proof overview of our main technical lemma. As
mentioned, we shall show that if one-way functions do not exist,
then Babai-Moran's round reduction method actually works. Towards this
we will rely on two tools:
\begin{itemize}
\item \emph{Pre-image sampling}. By the result of Impagliazzo and
  Levin \cite{ImpagliazzoL90}, the existence of so-called ``distributional one-way
  functions'' (function for which it is hard to sample a uniform
  pre-image) imply the existence of one-way function. So if one-way
  functions do not exist, we have that for every efficient function
  $f$, given a sample $f(x)$ for a random input $x$, we can
  efficiently sample a (close to random) pre-image $x'$.
\item \emph{Raz's sampling lemma} (from the literature on parallel
  repetition for 2-prover games and interactive arguments
\cite{Raz98,HastadPWP10, ChungP15}). This lemma states that if we sample $\ell$
uniform $n$-bit random variables $R_1, R_2, \ldots R_\ell$
conditioned on some event $W$ that happens with sufficiently large probability $\eps$,
then the conditional distribution $R_i$ of a randomly selected index $i$ will be
close to uniform. More precisely, the statistical distance will be
$\sqrt{\frac{\log(\frac{1}{\epsilon})}{\ell}}$, so even if $\eps$ is
tiny, as long as we have sufficiently many repetitions $\ell$, the
distance will be small.\footnote{Earlier works \cite{HastadPWP10,ChungP15} always
  used Raz' lemma when $\epsilon$ was non-negligible. In contrast, we
  will here use it also when $\epsilon$ is actually negligible.}
\end{itemize}
To see how we will use these tools, let us first recall the BM
transformation (and its proof for the case of
information-theoretically sound protocols).
To simplify our discussion, we here focus on showing how to collapse
a 3-round public-coin protocol between a prover $P$ and a public-coin
verifier $V$ into a 2-round protocol. We denote a transcript of the
3-round protocol $(p_1,r_1,p_2)$ where $p_1$ and $p_2$ are the prover
messages and $r_1$ is the randomness of the verifier. Let $n = |p_1|$
be the length of the prover message. The BM transformation collapses
this protocol into a 2-round protocol in the following two steps:
\begin{description}
\item[Step 1: Reducing soundness error:] First, use a form of parallel
  repetition to make the soundness error $2^{-n^2}$ (i.e.,
  \emph{extremely small}). More precisely,
  consider a 3-round protocol where $P$ first still send just $p_1$, next
  the verifier picks $\ell = n^2$ random strings
  $\vec{r} = (r_1^1,\ldots,r_1^{\ell})$, and finally $P$ needs to provide
  accepting answers $\vec{p_2}= (p_2^1, \ldots, p_2^{\ell})$ to all of the queries
  $\vec{r}$ (so that for every $i \in [\ell]$, $(p_1,
  r^i_1, p_2^i)$ is accepting transcript).
\item[Step 2: Swap order of messages:] Once the soundness error is
  small, yet the length of the first message is short, we can simply
  allow the prover to pick it first message $p_1$ after having
  $\vec{r}$. In other words, we now have a 2-round protocol where $V$
  first picks $\vec{r}$, then the prover responds by sending $p_1,
  \vec{p_2}$. This swapping preserves soundness by a simple union
  bound: since (by soundness) for every string $p_1$, the probability over $\vec{r}$
  that there exists some accepting response $\vec{r}$ is $2^{-n^2}$,
  it follows that with probability at most $2^n \times
  2^{-n^2} = 2^{-n}$ over $\vec{r}$, \emph{there exists some $p_1$} that has
  an accepting $\vec{p_2}$ (as the number of possible first messages
  $p_1$ is $2^n$). Thus soundness still holds (with a $2^n$
  degradation) if we allow $P$ to choose $p_1$ after seeing $\vec{r}$.
\end{description}
For the case of computationally sound protocols, the ``logic'' behind
both steps fail: (1) it is not
known how to use parallel repetition to reduce soundness error beyond being
negligible, (2) the union bound cannot be applied since, for
computationally sound protocols, it is not the case that responses
$\vec{p_2}$ do not exist, rather, they are just hard to find.
Yet, as we shall see, using the above tools, we present a different
proof strategy. More precisely, to capture computational hardness, we
show a reduction from any polynomial-time attacker $A$ that breaks soundness of the
collapsed protocol with some inverse polynomial probability $\epsilon$, to a polynomial-time attacker $B$ that breaks
soundness of the original 3-round protocol.

$B$ starts by sampling a random string $\vec{r'}$ and computes $A$'s
response given this challenge $(p'_1, \vec{p'_2}) \leftarrow
A(\vec{r'})$. If the response is not an accepting transcript, simply abort;
otherwise, take $p'_1$ and forward externally as $B$'s first message.
(Since $A$ is successful in breaking soundness, we have that $B$ won't
abort with probability $\epsilon$.)
Next, $B$ gets a verifier challenge $r$ from the external
verifier and needs to figure out how to provide an answer to it. If $B$
is lucky and $r$ is one of the challenges $r'_i$ in $\vec{r'}$, then $B$
could provide the appropriate $p_2$ message, but this unfortunately will only
happen with negligible probability. Rather, $B$ will try to get $A$ to
produce another accepting transcript $(p''_1, \vec{r''}, \vec{p''_2})$ that (1) still contains $p'_1$ as
the prover's first message (i.e., $p''_1 = p'_1$), and (2) contains $r$
in some coordinate $i$ of $\vec{r''}$. To do this,
$B$ will consider the function $f(\vec{r},z, i)$---which runs
$(p_1,\vec{p_2}) \leftarrow A(\vec{r};z)$ (i.e., $A$ has its
randomness fixed to $z$) and outputs $(p_1, r_i)$
if $(p_1,\vec{r}, \vec{p_2})$ is accepting and $\bot$ otherwise---and
runs the pre-image sampler for this function $f$ on $(p'_1,r)$ to recover
some new verifier challenge, randomness, index tuple $(\vec{r''},z,i)$ which leads $A(\vec{r''};z)$ to
produce a transcript $(p'_1,\vec{r''}, \vec{p''_2})$ of the desired form,
and $B$ can subsequently forward externally the $i$'th coordinate of
$\vec{p''_2}$ as its response and convince the
external verifier.

So, as long as the pre-image sampler indeed succeeds with high enough probability, we have managed
to break soundness of the original 3-round protocol. The problem is that the pre-image sampler is only required to work given
outputs that are correctly distributed over the range of the function
$f$, and the input $(p_1, r)$ that we now feed it may not be so---for
instance, perhaps $A(\vec{r})$ chooses the string $p_1$ as a function
of $\vec{r}$. So, whereas the marginal distribution of both $p_1$ and
$r$ are correct, the \emph{joint} distribution is not. In particular,
the distribution of $r$ conditioned on $p_1$ may be off. 
We, however, show how to use Raz's lemma to argue that
if the number of repetitions $\ell$ is sufficiently bigger
than the length of $p_1$, the conditional distribution of $r$ cannot
be too far off from being uniform (and thus the pre-image sampler
will work). On a high-level, we proceed as follows: 
\begin{itemize}
\item Note that in the one-way function experiment, we can
  think of the output distribution $(p_1, r)$ of $f$ on a random
  input, as having been produced by first
  sampling $p_1$ and next, if $p_1 \neq \bot$, sampling $\vec{r}$ conditioned on
  the event $W_{p_1}$ that $A$ generates a successful transcript with first-round prover message
  $p_1$, and finally sampling a random index $i$ and outputting $p_1$ and $r_i$ (and otherwise output $\bot$). 
\item Note that by an averaging argument, we have that with
  probability at least $\frac{\epsilon}{2}$ over the choice of $p_1$, $\Pr[W_{p_1}]
  \geq \frac{\epsilon}{2^{n+1}}$ (otherwise, the probability that
  $A$ succeeds would need to be smaller than $\frac{\epsilon}{2} + 2^n \times
  \frac{\epsilon}{2^{n+1}} = \epsilon$, which is a contradiction).
\item Thus, whenever we pick such a ``good'' $p_1$ (i.e., a $p_1$ such
  that $\Pr[W_{p_1}]
  \geq \frac{\epsilon}{2^{n+1}}$), by Raz' lemma the
distribution of $r_i$ for a random $i$ can be made $\frac{1}{p(n)}$
close to uniform for any polynomial $p$ by choosing $\ell$ to be
sufficiently large (yet polynomial). Note that even though the lower bound on $\Pr[W_{p_1}]$ is
negligible, the key point is that it is independent of $\ell$ and
as such we can still rely on Raz lemma by choosing a sufficiently
large $\ell$. (As we pointed out above, this usage of Raz' lemma even
on very ``rare'' events---with negligible probability mass---is different
from how it was previously applied to
argue soundness for computationally sound protocols \cite{HastadPWP10, ChungP15}.)
\item It follows that conditioned on picking such a ``good'' $p_1$,
  the pre-image sampler will also successfully generate correctly
  distributed preimages if we feed him $p_1,r$ where $r$ is randomly
  sampled. But this is exactly the distribution that $B$ feeds to the
  pre-image sampler, so we conclude that with probability
  $\frac{\epsilon}{2}$ over the choice of $p_1$, $B$ will manage to
  convince the outside verifier with probability close to 1.
\end{itemize}
This concludes the proof overview for 3-round protocols. When the
protocol has more than 3 rounds, we can apply a similar method to collapse
the last rounds of the protocol. The analysis now needs to be
appropriately modified to condition also on the prefix of the partial
execution up until the last rounds.

\section{Preliminaries}
We assume familiarity with basic concepts such as Turing machines,
interactive Turing machine,
polynomial-time algorithms, 
probabilistic polynomial-time algorithms ($\ppt$), non-uniform
polynomial-time and non-uniform $\PPT$ algorithms.
A function $\mu$ is said to be \emph{negligible} if for every
polynomial $p(\cdot)$ there exists some $n_0$ such that for all $n >
n_0$, $\mu(n) \leq \frac{1}{p(n)}$.
For any two random variables $X$ and $Y$, we let 
$\mathsf{SD}(X,Y) = \max_{T \subseteq U}|\Pr[X \in T]-\Pr[Y\in T]|$ denote the
\emph{statistical distance} between $X$ and $Y$.
\remove{
\BP\label{prop:KLstat}
For any two random variables $X$ and $Y$ we have that
$$\KL(X||Y) \geq 2 ||X-Y||^2$$
\EP}

\paragraph{Basic Complexity Classes} 
Recall that $\P$ is the class of languages $L$ decidable in
polynomial time (i.e., there exists a polynomial-time algorithm $M$
such that for every $x\in \{0,1\}^*$, $M(x) = L(x)$), $\Ppoly$ is the
class of languages decidable in non-uniform polynomial time, and
$\BPP$ is the class of languages decidable in probabilistic
polynomial time with probability $2/3$ (i.e., there exists a $\PPT$ $M$
such that for every $x\in \{0,1\}^*$, $\Pr[M(x) =L(x)]> 2/3$ where we abuse
of notation and define $L(x) = 1$ if $x\in L$ and $L(x) = 0$
otherwise.)

We refer to a relation $\cR$ over pairs $(x,y)$ as being
\emph{polynomially bounded} if there exists a polynomial $p(\cdot)$
such that for every $(x,y) \in \cR$, $|y| \leq p(|x|)$.
We denote by $L_{\cR}$ the language characterized by the ``witness relation''
$\cR$---i.e., $x \in L$ iff there exists some $y$ such that $(x,y) \in \cR$.
We say that a
relation $\cR$ is \emph{polynomial-time} (resp. non-uniform
polynomial-time) if $\cR$ is
polynomially-bounded and the languages consisting of pairs $(x,y) \in
\cR$ is in $\P$ (resp. $\Ppoly$).
$\NP$ (resp $\NPpoly$) is the class of languages $L$
for which there exists a polynomial-time (resp. non-uniform
polynomial-time) relation $\cR$ such that $x
\in L$ iff there exists some $y$ such that $(x,y) \in \cR$.

\paragraph{Search Problems}
A search problem $\cR$ is simply a polynomially-bounded relation; an
$\NP$ search problem $\cR$ is a polynomial-time relation. 
We say that the search problem is \emph{solvable in polynomial-time
(resp. non-uniform polynomial time)} if there
exists a polynomial-time (resp. non-uniform polynomial-time) algorithm $M$ that for every $x
\in L_{\cR}$ outputs a ``witness'' $y$ such that $(x,y) \in
\cR$. Analogously, $\cR$ is \emph{solvable in $\PPT$} if there exists some $\PPT$
$M$ that for every $x
\in L_{\cR}$ outputs a ``witness'' $y$ such that $(x,y) \in
\cR$ with probability $2/3$.

An $\NP$ search problem $\cR$ is \emph{total} if for every $x \in \{0,1\}^*$ there exists some $y$ such
that $(x,y) \in \cR)$ (i.e., every instance has a witness). We refer
to $\FNP$ (function \NP) as the class of $\NP$ search problems and 
$\TFNP$ (total-function $\NP$) as the class of total $\NP$ search
problems.

\subsection{One-way functions}
We recall the definition of one-way functions (see e.g., \cite{Gol01}). Roughly speaking, a
function $f$ is one-way if it is polynomial-time computable, but hard to
invert for $\PPT$ attackers. The standard (cryptographic) definition of a one-way function
requires every $\PPT$ attacker to fail (with high probability) on all
sufficiently large input lengths. We will
also consider a weaker notion of an \emph{infinitely-often} one-way
function \cite{OstrovskyW93} which only requires the $\PPT$ attacker to fail for
infinitely many inputs length (in other words, there is no $\PPT$
attacker that succeeds on all sufficiently large input lengths, analogously to
complexity-theoretic notions of hardness).

\BD\label{def:owf} Let $f: \bitset^* \rightarrow \bitset^*$ be a polynomial-time
computable function. $f$ is said to be a \emph{one-way function (OWF)} if for every $\ppt$
algorithm $A$, there exists a negligible function $\mu$ such that for
all $n \in \NN$,
	$$ \Pr[x \leftarrow \bitset^n; y = f(x) : A(1^n,y) \in f^{-1}(f(x)) ] \leq \mu(n) $$
        
        $f$ is said to be an \emph{infinitely-often one-way function (ioOWF)} if the above
condition holds for infinitely many $n\in\NN$ (as opposed to all). 
        \ED
        We may also consider a notion of a
        \emph{non-uniform} (a.k.a. ``auxiliary-input'') one way function, which is identically
        defined except that (a) we allow $f$ to be computable by a
        non-uniform $\PPT$, and (b) the attacker $A$ is also allowed to be a
        non-uniform  $\PPT$.        
\remove{

\BD [Efficient Family of Pairwise Independent Hash Functions] Let $\cal H$ be a family of functions where each function $h \in {\cal H}$ goes from $\bitset^m$ to $\bitset^n$. We say that 
$\cal H$ is a an efficient family of pairwise independent hash functions if (i) the functions $h \in {\cal H}$ can be described with a polynomial (in $n$) number of bits; (ii) there is 
a polynomial (in $n$) time algorithm to compute $h \in {\cal H}$; (iii) for all $x \neq x' \in \bitset^m$ and for all $y,y' \in \bitset^n$ 
$$ \Pr[h \leftarrow \cH:h(x)=y \; \mbox{ and } \; h(x')=y'] = 2^{-2n} $$
\ED
}

\subsection{Interactive Proofs and Arguments}
We recall basic definitions of interactive proofs \cite{GMR89,BM88} and
arguments \cite{BrassardCC88}. An
interactive protocol $(P,V)$ is a pair of interactive Turing machine;
we denote by $\inter{P_1,P_2}(x)$ the
output of $P_2$ in an interaction between $P_1$ and $P_2$ on common input $x$.
\BD An interactive protocol $(P,V)$ is an \emph{interactive
  proof system} for a language $L \subseteq \{0,1\}^*$, if $V$ is
$\PPT$ and the following conditions hold:
\BDE
\item[Completeness:] There exists a negligible function $\mu(c\dot)$ such that for every $x \in L$, 
  $$\Pr[\inter{P,V}(x)=1] \geq 1-\mu(|x|)$$
\item[Soundness:] For every Turing machine $P^*$, there exists a
  negligible function $\mu(\cdot)$ such that for every $x \not\in L$, 
  $$\Pr[\inter{P^*,V}(x)=1] \leq \mu(|x|)$$
\EDE
If the soundness condition is relaxed to only hold for all non-uniform
$\PPT$ $P^*$, we refer to $(P,V)$ as an \emph{interactive argument}
for $L$. We refer to $(P,V)$ as a \emph{public-coin proof/argument system} if $V$ simply sends the
outcomes of its coin tosses to the prover (and only performs
computation to determine its final verdict).

Whenever $L \in \NP$, we say that
$(P,V)$ has an \emph{efficient prover} if there exists some witness
relation $\cR$ that characterizes $L$ (i.e., $L_{\cR} = L$) and a \ppt\ $\widetilde{P}$ such
that $P(x) = \widetilde{P}(x,w)$ satisfies the completeness condition for every
$(x,w) \in \cR$.
\ED

\subsection{Average-Case Complexity}\label{sec:heur}
We recall some basic notions from average-case complexity. 
A \emph{distributional problem} is a pair $(L,\cD)$ where $L \subseteq
\bitset^*$ and $\cD$ is a $\PPT$; we say that $(L,\cD)$ is an $\NP$
\ifma{(resp. $\NPpoly$)} distributional problem if $L \in \NP$ (resp. $L \in
\NPpoly)$. Roughly speaking, a distributional
problem $(L,\cD)$ is hard-on-average if there does not exist some $\ppt$ 
algorithm that can decide instances drawn from $\cD$ with probability
significantly better than $1/2$.

\BD [$\delta$-hard-on-the-average]
We say that a distributional problem $(L,\cD)$ is
\emph{$\delta$-hard-on-the-average ($\delta$-HOA)} if there does not exist
some $\ppt$ $A$ such that for every sufficiently large $n \in \NN$,
$$\Pr[x \la \cD(1^n): A(1^n,x) = L(x)] >  1 - \delta$$
We say that a distributional problem $(L,\cD)$ is simply
 \emph{hard-on-the-average (HOA)} if it is $\delta$-HOA for some $\delta > 0$.
\ED
We also define an notion of HOA w.r.t. non-uniform $\PPT$ algorithm
(\emph{nuHAO}) in exactly the same way but where we allow $A$ to be a
non-uniform $\PPT$ (as opposed to just a $\PPT$.

The above notion average-case hardness (traditionally used in the
complexity-theory literature) is defined analogously to the notion of
an \emph{infinitely-often} one-way function: we simply require every $\PPT$
``decider'' to fail for infinitely many $n\in \NN$. For our purposes, we
will also rely on an ``almost-everywhere'' notion of average-case hardness
(similar to standard definitions in the cryptography, and analogously
to the definition of a one-way function), where we require that every
decider fails on \emph{all} (sufficiently large) input lengths.

\BD [almost-everywhere hard-on-the-average (aeHOA)]
We say that a distributional problem $(L,\cD)$ is
\emph{almost-everywhere $\delta$ hard-on-the-average ($\delta$-aeHOA)}
if there does not exist some $\ppt$ $A$ such that for infinitely many $n \in \NN$,
$$\Pr[x \la \cD(1^n): A(1^n,x) = L(x)] > 1-\delta$$
We say $(L,\cD)$ is \emph{almost-everywhere hard-on-the-average
  (aeHOA)} if $(L,\cD)$ is $\delta$-aeHOA for some $\delta > 0$.
\ED

We move on to defining hard-on-the-average \emph{search problems}. 
A \emph{distributional search problem} is a pair $(\cR,\cD)$ where
$\cR$ is a search problem and $\cD$ is a $\PPT$. If
$\cR$ is an $\NP$ search problem (resp. $\NPpoly$ search problem), we
refer to $(\cR,\cD)$ as an distributional $\NP$  (resp. $\NPpoly$) search problem.

Finally, we say that a distributional search problem $(\cR,\cD)$
is \emph{\promise} if for every $n$ and every $x$ in the support
of $\cD(1^n)$, it holds that $x \in L_{\cR}$. (That is, $\cD$ only
samples true instances.)

\BD [hard-on-the-average search (SearchHOA)]
We say that a distributional search problem $(\cR,\cD)$ is
\emph{$\delta$-hard-on-the-average ($\delta$-SearchHOA)} if there does
not exist some $\PPT$ $A$ such that for every sufficiently large $n \in N$,
$$\Pr[x \la \cD(1^n); (w,x) \leftarrow A(1^n,x) : \left((L_{\cR} (x) = 1) \Rightarrow (x,w) \in \cR\right)] > 1 - \delta$$
$(\cR,\cD)$ is simply \emph{SearchHOA} if there exists $\delta >0$ 
such that $(\cR,\cD)$ is $\delta$-SearchHOA. 
\ED
We can analogously define an almost-everywhere notion,
\emph{aeSearchHOA}, of SearchHAO (by replacing ``for every
sufficiently large $n\in N$"
with ``for infinitely many $n \in N$'') as well as a non-uniform
notion, \emph{nuSearchHOA}, (by replacing $\PPT$ with non-uniform
$\PPT$).

The following lemmas which essentially directly follow from the result
of \cite{ImpagliazzoL90, Ben-DavidCGL92,Trevisan05} will be useful
to us. (These results were originally only stated for the standard
notion of HOA, whereas we will require it also for the
almost-everywhere notion; as we explain in more detail in Appendix
\ref{app:extendedavg}, these results however directly apply also for the
almost-everywhere notion of HOA.)
The first results from \cite{ImpagliazzoL90} (combined with
\cite{Trevisan05}) shows that without loss
of generality, we can restrict our attention to the uniform
distribution over statements $x$; we denote by $\cU_p$ a $\PPT$ such
that $\cU_p(1^n)$ simply samples a random string in $\{0,1\}^{p(n)}$.

\BL[Private to public distributions]\label{lem:privatetopublic}
Suppose there exists a distributional $\NP$ problem $(L,\cD)$
that is HOA (resp., aeHOA or nuHAO). Then, there exists a
polynomial $p$ and an $\NP$-language $L'$ such that $(L',\cU_p)$ is
HAO (resp. aeHOA or nuHOA).
\EL
The next result from \cite{Trevisan05}) shows that when the
distribution over instances is uniform, we can amplify the hardness.
\BL[Hardness amplification]\label{lem:anydelta}
Let $p$ be a polynomial and suppose there exists a distributional $\NP$-problem $(L,\cU_p)$ that is
HOA (resp., aeHOA or nuHOA). Then, for every $\delta < \frac{1}{2}$, 
there exists some polynomial $p'$ and $\NP$ language $L'$ such that
$(L',\cU_{p'})$ is $\delta$-HOA  (resp., $\delta$-aeHOA or $\delta$-nuHOA).
\EL 
Finally, by \cite{Ben-DavidCGL92} (combined with
\cite{Trevisan05}, \cite{ImpagliazzoL90}) we have a decision-to-search reduction.
\BL[Search to decision]\label{lem:searchtodecision}
Suppose there exists a distributional $\NP$ (resp. $\NPpoly$) search problem
$(\cR,\cD)$ that is SearchHOA (resp., nuSearchHOA). Then, there a
polynomial $p$ and an $\NP$ (resp. $\NPpoly$) language $L$ such that
$(L',\cU_p)$ is HOA (resp., nuHOA).
\EL





\section{Interactive Puzzles}
Roughly speaking, an \emph{interactive 
  puzzle} is 
described by an interactive polynomial-time challenger $\Chal$
having the property that
(a) there exists an inefficient $\Adv$ that succeeds in convincing $\Chal(1^n)$ with probability negligibly
close to 1, yet (b) no $\PPT$ attacker $\Adv^*$ can make $\Chal(1^n)$ output 1 with inverse polynomial
probability for sufficiently large $n$. 

\BD[interactive puzzle]\label{def:iopuz} 
An interactive algorithm $\Chal$ 
is referred to as a
{\em $k(\cdot)$-round puzzle} if the following conditions hold:
\BDE
\item [$k(\cdot)$-round, publicly-verifiability:] $\Chal$ is an (interactive)
  $\PPT$ that on input $1^n$ (a) only communicates in $k(n)$ communication
  rounds,
  and (b) only performs some deterministic 
computation as a function of the transcript to determine its final verdict.
\item[Completeness/Non-triviality:] There exists a (possibly
  unbounded) Turing machine $\Adv$ and a negligible function $\mu_C(\cdot)$ such
  that for all $n \in \NN$,
$$ \Pr[\inter{\Adv,\Chal}(1^n)=1] \geq 1-\mu(n)$$
\item[Computational Soundness:] There does not exist a $\ppt$ machine $\Adv^*$ and
  polynomial $p(\cdot)$ such that for all sufficiently large $n\in \NN$, 
  $$ \Pr[\inter{\Adv^*,\Chal}(1^n)=1] \geq \frac{1}{p(n)}$$
\EDE
\ED
In other words, a $k(\cdot)$-round puzzle, $\myprot{\Adv}{\Chal}$,
gives rise to an $k(\cdot)$-round interactive proof $(P,V)$ (where $P =
\Adv, V = \Chal$) for the ``trivial'' language $L = \{0,1\}^*$ with
the property that there does not exist a $\PPT$ prover that succeeds
in convincing the verifier with inverse polynomial probability for
all sufficiently large $n$.

We will consider several restricted, or alternative, types of puzzle:
\BI
\item We refer to the puzzle $\Chal$ as being
\emph{public-coin} if $\Chal$ simply sends the
outcomes of its coin tosses 
in each communication round.
\item We may also define an \emph{almost-everywhere} notion of a puzzle by
replacing ``for all sufficiently large $n \in \NN$'' in the soundness condition with ``for
infinitely many $n \in \NN$'', and a \emph{non-uniform} notion of a
puzzle $\Chal$ which allows both $\Chal$ and $\Adv^*$ to be \emph{non-uniform}
$\PPT$ (as opposed to just $\PPT$).
\item Finally, a puzzle $\myprot{\Adv}{\Chal}$ is said to have \emph{perfect completeness} if
the ``completeness error'', $\mu_C(n)$, is 0---in other words,
the completeness condition holds with probability 1.
\EI

\BR\label{rem:par} One can consider a more relaxed notion of a
$(c(\cdot),s(\cdot))$-puzzle for $c(n)>s(n)+\frac{1}{\poly(n)}$, where
the completeness condition is required to hold
with probability $c(\cdot)$ for every sufficiently large $n \in \NN$, and the soundness condition holds with
probability $s(\cdot)$ for every sufficiently large $n \in \NN$. But, by ``Chernoff-type'' parallel-repetition theorems
for computationally-sound protocols \cite{PassV12,Haitner,HastadPWP10,ChungL10,ChungP15}, the
existence of such a $k(\cdot)$-round $(c(\cdot),s(\cdot))$-puzzle implies the
existence of a $k(\cdot)$-round puzzle. The same holds for
almost-everywhere (resp. non-uniform) puzzles.
\ER
\noindent

\subsection{Characterizing 2-round Public-coin Puzzles}
In this section we make some basic observations regarding 2-round
public-coin puzzles; these results mostly follow using standard results in the
literature.
We begin by observing that the existence of ioOWF imply the
existence of 2-round public-coin puzzles.
\BP\label{lem:2puzowf}
Assume the existence of ioOWFs (resp. non-uniform ioOWF). Then, there
exists a $2$-round public-coin puzzle (reps. non-uniform puzzle).  
\EP
\BPF
By the result of Rompel \cite{Rompel90} (see also \cite{KK05,HHRVW10}), we have
that ioOWFs imply the existence of infinitely-often
``second-preimage'' resistant hash-function families that compress $n$
bits to $n/2$ bits.\footnote{Roughly
  speaking, a family of public coin hashfunctions $H$ having the
  property that for a random $h \in H$ and random input $x$, it is hard for
  any $\PPT$ to find a different $x'$ of the same length that collides with $x$ under $h$
  (that is, $h(x) = h(x')$, $|x| = |x'|$ yet $x\neq x'$. Rompel's
  theorem was only stated for standard OWFs (as opposed
  to ioOWFs, but the construction and proof directly also works for
  the infinitely-often variant as well.} This, in turn,
directly yields a simple 2-round puzzle where the challenger $\Chal(1^n)$
uniformly samples a hashfunction $h$ and input $x \in \{0,1\}^n$ and sends $(h,x)$
to the adversary; $\Chal$ accepts a response $x'$ if $|x'| = n$, $x'
\neq x$ and $h(x') = h(x)$. Since the hash function is compressing, we have that there exists a negligible function $\mu$
such that with probability $1-\mu(n)$, a random $x \in \{0,1\}^n$
will have a ``collision'' $x'$ and thus an unbounded $\Adv$ can easily
find a collision and thus completeness
follows. Computational soundness, on the other hand, directly from the
(infinitely-often) second-preimage resistance property. The same result
holds also if we start with non-uniform ioOWFs, except that we now get
a non-uniform puzzle.
\EPF
We turn to showing that any aeHOA distributional $\NP$ problem implies a
2-round puzzle. (In fact, it even implies an almost-everywhere puzzle.)
\BL\label{lem:hoa2puz}
Suppose there exists a distributional $\NP$ problem $(L,\cD)$ that is
aeHOA. Then there exist an (almost-everywhere) $2$-round public-coin puzzle.
\EL
\BPF Assume there exists a distributional problem $(L,\cD)$ such that
$L \in \NP$ and $(L,D)$ is aeHOA. From  Lemma~\ref{lem:anydelta} and
Lemma~\ref{lem:privatetopublic}, we can conclude that there exists a
polynomial $p$ and a distributional $\NP$ problem $(L',\cU_p)$ that is
$\delta$-aeHOA for $\delta = \frac{3}{8}$. Let $\cR'$ be some $\NP$ relation corresponding to $L'$. 
Consider a puzzle $\myprot{\Adv}{\Chal}$ where $\Chal(1^n)$
samples a random $x\in \{0,1\}^{p(n)}$ and accepts a response $y$ if $(x,y)
\in \cR'$. We will show that $\myprot{\Adv}{\Chal}$ is a
$(\frac{3}{8}, \frac{1}{4})$-puzzle which by Remark~\ref{rem:par}
implies the existence of a 2-round almost-everywhere puzzle.
To show completeness, consider an inefficient algorithm $\Adv$ that on input $(1^n,x)$ tries to find a witness $y$ (using
brute-force) such that $(x,y)
\in \cR'$ and if it is successful sends it to $\Chal$ (and otherwise simply aborts).  
Observe that for all sufficiently large $n \in \NN$,  for a random $x
\leftarrow \{0,1\}^{p(n)}$ we have that $\Pr[x \in L'] > \frac{3}{8}$; otherwise, $(L',\cU_p)$ can be decided with probability $1-\frac{3}{8}$
for infinitely many $n \in \NN$ contradicting its
$\frac{3}{8}$-aeHOA property.\footnote{Note
  that this is where were are crucially relying on the
  almost-everywhere hardness of the distributional problem.}
It follows that for all sufficiently large $n \in N$, $\Adv$ convinces
$\Chal$ with probability $\frac{3}{8}$ and thus completeness of
$\myprot{\Adv}{\Chal}$ follows.

To prove soundness, assume for contradiction that there exists a  $\ppt$
algorithm $\Adv^*$  such that  $\defsuc{\Adv^*}{n} > \frac{1}{4}$ for
infinitely many $n$. Consider the machine $M(x)$ that runs $\Adv^*(1^{|x|},
x)$ and outputs 1 if $\Adv^*$ outputs a valid witness for $x$ and
otherwise outputs a random bit. By definition, $M$ solves the distributional problem $(L',\cU_p)$
with probability $> \frac{1}{4}+\frac{1}{2}\left(1 -
  \frac{1}{4}\right) = \frac{5}{8} = 1 - \frac{3}{8}$ for
infinitely many $n$, which contradicts the $\frac{3}{8}$-aeHAO
property of $(L',\cU_p)$.
\EPF

We now turn to showing that 2-round puzzles imply a HOA distributional
$\NP$ problem. It will be useful for the sequel to note that the same
result also holds in the non-uniform setting.
\BL\label{lem:puz2hoa}
 Suppose there exists a $2$-round public-coin puzzle (resp. a non-uniform puzzle). Then, there exists a
 distributional $\NP$ problem (resp. distributional $\NPpoly$ problem)
 that is HOA (resp. nuHOA).
 \EL
\BPF Let $\Chal$ be a 2-round public-coin puzzle (resp. 2-round non-uniform
puzzle). Let $\ell(\cdot)$ be an upper bound on the amount of
randomness used by $\Chal$. Consider the
$\NP$-relation (resp. $\NPpoly$-relation) $\cR$ that includes all tuples $((pad,x),y)$ such that
$\Chal(1^{|pad|})$ given randomness $x\in \ell(|pad|)$ accepts upon receiving $y$, and the
sampler $\cD(1^n)$ that picks a random $x \in \bitset^{\ell(n)}$ and outputs
$(1^n,x)$.
We argue next that $(\cR,\cD)$ is $\frac{1}{8}$-SearchHOA (resp $\frac{3}{8}$-nuSearchHOA), which
concludes the proof by applying Lemma~\ref{lem:searchtodecision}. 
Assume for contradiction that there exists a $\PPT$ (resp. non-uniform
$\PPT$) machine $M$ that
solves $(\cR,\cD)$ with probability $> 1- \frac{1}{8} =
\frac{7}{8}$ for all $n> n_0$.
By the completeness of $\myprot{\Adv}{\Chal}$, there exists some
$\Adv, n_1$ such that such that for every $n>n_1$, $\defsuc{\Adv}{n} >
\frac{7}{8}$. This implies that for all $n > n_1$, for at most an
$\frac{1}{8}$ fraction of $\ell(n)$-bit strings $x$, $(1^n,x) \notin L_{\cR}$). In particular, for every $n > \max(n_0,n_1)$, for a random $x \in \{0,1\}^{\ell(n)}$, $M(1^n,x)$ must output a valid witness $y$
for $x$ with probability $> \frac{7}{8}-\frac{1}{8} > \frac{1}{2}$,
and can thus be used to break the soundness of the puzzle with
probability $> \frac{1}{2}$ for all sufficiently large $n$ which is a contradiction.
\EPF

If the 2-round puzzle has perfect completeness, essentially the same
proof gives a SearchHOA problem in $\TFNP$ as the relation $\cR$
constructed in the proof of Lemma \ref{lem:puz2hoa} is total if
the puzzle has perfect completeness.
\BL\label{lem:ioperfpuzTFNP}
Suppose there exists a $2$-round public-coin puzzle (resp. almost-everywhere puzzle) with perfect
completeness. Then, there exists some search problem $\cR \in \TFNP$
and some $\PPT$ $\cD$ such that the distributional search problem 
$(\cR,\cD)$ is SearchHAO (aeSearchHAO).
\EL

\section{The Round-Collapse Theorem}
In this section, we prove our main technical lemma---a
round-collapse theorem for $O(1)$-round puzzles. 

\subsection{An Efficient Babai-Moran Theorem}
Our main lemma shows that if ioOWF do not exist, the the Babai-Moran
transformation preserves computational soundness.
\BL\label{lem:collapse}
Assume there exists a $k(\cdot)$-round 
public-coin puzzle such that $k(n)\geq 3$.
Then, either there exists an ioOWF, or there exists a $(k(\cdot)-1)$-round 
public-coin puzzle. Moreover, if the $k(\cdot)$-round puzzle has perfect
completeness, then either there exists an ioOWF, or a
$(k(\cdot)-1)$-round public-coin puzzle with perfect-completeness. 
\EL
\BPF Consider some $k(\cdot)$-round public-coin puzzle $\myprot{\Adv}{\Chal}$ and assume for
contradiction that ioOWF do not exist. We will show that
Babai-Moran's (BM) \cite{BM88} round
reduction works in this setting and thus we can obtain a
$(k(\cdot)-1)$-round puzzle.

Note that if ioOWF do not
exist, every polynomial-time computable function is
``invertible'' with inverse polynomial probability for all
sufficiently long input
lengths $n$. In fact, since by 
\cite{ImpagliazzoL90}, the existence of \emph{distributional one-way
  functions} implies the existence of one-way functions (and this
results also works in the infinitely-often setting), we can conclude
that if ioOWF do not exist, for any polynomial $p(\cdot)$, and any
polynomial-time computable function $f$, there exists a $\ppt$
algorithm $\Inv$ such that, for sufficiently large $n$, the following
distributions are $\frac{1}{p(n)}$-statistically close. 
\BI
\item $\{x \la \bitset^n:(x,f(x))\}$
\item $\{x \la \bitset^n; y = f(x): (\Inv(y),y))\}$
\EI
In this case, we will say that $\Inv$ inverts $f$ \emph{with
  $\frac{1}{p(n)}$-statistical closeness}.
We now proceed to show how to use such an inverter to prove that BM's round-collapse
transformation works on $\myprot{\Adv}{\Chal}$. To simplify notation, we will make the following assumptions that are without loss of generality:
\BI
\item  $\myprot{\Adv}{\Chal}$ has at least 4 communication rounds and $\Chal$
sends the first message; we can always add an initial dummy message to
achieve this, while only increasing the number of round by 1. We will
then construct a new puzzle that has $k(\cdot)-2$ rounds (which
concludes the theorem). Since, in any puzzle, $\Adv$ sends the final message, this implies we can assume $k(\cdot)$ is even. To make our notations easier to read, we show how to reduce a $2k(\cdot)$-round protocol to a $2k(\cdot)-2$ rounds. 

\item There exists
polynomials $\ell_c,\ell_a$ such that all
messages from $\Chal(1^n)$ are of (the same) length $\ell_c(n)$ and all the
messages from $\Adv(1^n)$ need to be of length $\ell_a(n)$ (or else
$\Chal$ rejects). Furthemore, $\ell_a(\cdot)$ and $\ell_c(\cdot)$ are
polynomial-time computable, and strictly increasing.
\EI
We denote by $\Chal(1^n,p_1,p_2,\ldots,p_{i};r_{\Chal})$ the
$(i+1)^{st}$-message (i.e., the message to be sent in round $2i+1$
round) from $\Chal$ where $r_\Chal$ is $\Chal$'s randomness
randomness and  $p_1,p_2,\ldots,p_{i}$ are bit strings (representing
the messages received from $\Adv$ in the first $2i$ rounds). %
\newcommand{\reps}{m}
\newcommand{\rta}{h}
Let $\reps(n)$ be $(\ell_a(n)+4)(\log(n))^2$ rounded upwards to the next power
of two.\footnote{We round to the next power of 2 to make it easy to
  sample a random number in $[\reps(n)]$; this is just to simplify presentation/analysis} We will show that the BM
transformation works (if ioOWF do not exist), when using $\reps(\cdot)$
repetitions.  
More precisely, consider the following $(2k(\cdot)-2)$-round puzzle challenger $\myprot{\np}{\nv}$ that on input $(1^n,p_1,\ldots,p_{i};r_{\nv})$ proceeds as  follows: 
\begin{enumerate}
\item If $i < k(n)-2$, output $r_{i+1}$ (i.e., proceed just like
  $\Chal$ before round $2k(n)-3$);  
\item If $i = k(n)-2$, output
  $(r_{k(n)-1},r^1_{k(n)},\ldots,r^{\reps(n)}_{k(n)})$ (i.e., in round
  $2k(n)-3$, send the
  original challenge for round $2k(n)-3$ as well as a
  ``$m(n)$-wise parallel-repetition'' challenge for the original round $2k(n)-1$);
\item If $i = k(n)-1$ (i.e., after receiving the message in the last
  round), output 1 if and only if
  $$\Chal(1^n,p_1,p_2,\ldots,p_{k(n)-1},p_{k(n)}^i;r_1,r_2,\ldots,r_{k(n)-1},r^i_{k(n)})=1$$
  for every $i \in [\reps(n)]$ (i.e., all the parallel instances are accepting),
\end{enumerate}
 where $r_{\nv}$ is interpreted as $(r_1,r_2,\ldots,r_{k(n)-1},r^1_{k(n)},\ldots,r^{\reps(n)}_{k(n)})$

We will show that $\myprot{\np}{\nv}$ is a $(99/100, 1/2)$-puzzle, and thus
by Remark~\ref{rem:par} this implies a puzzle with the
same number of rounds. 

We first define some notation: 
\begin{itemize}
  \item Given a transcript $T =
(r_1,p_1,\ldots,p_{k(n)-2},r_{k(n)-1},r^1_{k(n)},\ldots,r^{\reps(n)}_{k(n)},p_{k(n)-1},p^1_{k(n)},\ldots,p^{\reps(n)}_{k(n)})$
of an interaction between $\nv$ and an adversary, 
we let $T_{\leq k-1} = (r_1,p_1,\ldots,p_{k(n)-2},r_{k(n)-1})$ denote the
transcript up to and including the round where $\Chal$ (in the
emulation done by $\nv$) sends it $(k(n)-1)$'st message.
\item We say that $T$ is \emph{accepting} if
$$\nv(p_1,\ldots,p_{k(n)-2}, p_{k(n)-1},p^1_{k(n)},\ldots,p^{\reps(n)}_{k(n)}; r_1, \ldots
r_{k(n)-1}, r^1_{k(n)},\ldots,r^{\reps(n)}_{k(n)}) = 1$$ (i.e,. if $\nv$ is accepting
in the transcript).
\end{itemize}

\paragraph{Completeness:} Completeness (in fact with all but negligible
probability) follows directly from original proof by Babai-Moran
\cite{BM88}.

\paragraph{Soundness:} Assume for contradiction that there exists a $\ppt$
algorithm $\Adv^*$ 
that convinces $\nv$ on common input $1^n$
with probability $\eps(n)$ such that $\epsilon(n) >\frac{1}{2}$ for
all sufficiently large $n$. Let $\rta(\cdot)$ be a polynomial such
that $\Adv^*$ runs in time at most $\rta(n)$ when its first input is
$1^n$. We assume without loss of generality that $A^*$ only sends a real
last message if $\nv$ will be accepting it (note that since $\nv$
is public coin, $A^*$ can verify this, so it is without loss of
generality), and otherwise sends $\bot$ as its last message. 

On a high-level, using $\Adv^*$
and the fact that polynomial-time computable functions are
``invertible'', we will construct a $\ppt$ $B$ such that
$\defsuc{B}{n} \geq \frac{1}{64}$ for sufficiently large $n$, which
contradicts the soundness of the original $2k(n)$-round puzzle $\Chal$. 
Towards constructing $B$, we first define a polynomial-time algorithm $M$ on which we
will apply the inverter $\Inv$. 
As described in the introduction, we will consider an algorithm $M$ that
operates on inputs of the form $(1^n, i,r_M)$ where $i$ is an index of
one of the $\reps(n)$ parallel sessions and $r_M$ contains the
randomness of $\Adv$ and $\nv$. To correctly parse such inputs, let
$\ell_M(n) = n+\log( \reps(n)) + \rta(n) +
\ell_c(n)\cdot(k(n)-1+\reps(n))$
and note that by our assumption on $\ell_c(n)$ and $\ell_a(n)$, this is a strictly
increasing and polynomial-time computable function.
In the rest of the proof, whenever the security parameter $n$ is clear from context, we omit it and let $k=k(n),\ell_c = \ell_c(n), \ell_a = \ell_a(n),\eps = \eps(n), \reps = \reps(n)$ and $\rta = \rta(n)$.
Now, consider the machine $M$ that on input $u$ internally incorporates the code of $\Adv^*$ and proceeds as follows:
\BE
\item $M$ finds an $n$ such that $\ell_M(n) = |u|$ (simply by
  enumerating different $n$ from 1 up to $|u|$). If no such $n$
  exists, then $M$ outputs $\bot$ and halts. Otherwise, $M$ interprets
  $u$ as  $(pad,i,r_M)$ such that  $|pad|=n$, $|i| =
  \log(\reps(n))$, $r_M=(z,r_1,r_2,\ldots,r_{k-1},r_{k}^1,\ldots,r_{k}^{\reps})$, $z \in \{0,1\}^{\rta}$, and all the strings $r_1,r_2,\ldots,r_{k-1},r_{k}^1,\ldots,r_{k}^\reps$ are in $\{0,1\}^{\ell_c}$.
\item It internally emulates an execution between $\Adv^*$ and $\nv$
  on common input $1^n$ and respectively using randomness $z$ and
  $(r_1,r_2,\ldots,r_{k-1},r_{k}^1,\ldots,r_{k}^{\reps})$. Let $$T
  =(r_1,p_1,\ldots,p_{k-2},r_{k-1},r^1_{k},\ldots,r^{\reps}_{k},p_{k-1},p^1_k,\ldots,p^\reps_k)$$
  denote the transcript of the interaction.
 \item If $T$ is accepting, then $M$ outputs 
$$(1^{|pad|},T_{\leq k-1}, p_{k-1},r^i_{k}),$$
and otherwise $\bot$.
\EE
Let $\Inv$ be an inverter for $M$ with $\frac{1}{n}$
statistical-closeness for all sufficiently large $n$---such an inverter
exists due to our assumption on the
non-existence of ioOWFs.  

\renewcommand{\widr}{s}
\renewcommand{\hatr}{t}
\renewcommand{\widp}{q}
\newcommand{\widi}{j}
We are now ready to describe our adversary $B$ for the $2k$-round
puzzle. 
$B$ on input $(1^n,r_1,r_2,\ldots,r_{i};r_B)$ proceeds as follows:
\BE
\item $B$ interprets $r_B$ as
  $(z,pad,\widr^1_{k-1},\ldots,\widr^{\reps}_{k-1})$ such that $z \in \{0,1\}^{\rta},pad \in \{0,1\}^n$ and all the strings $\widr^1_{k-1},\ldots,\widr^{\reps}_{k-1}$ are in $\{0,1\}^{\ell_c}$.
\item\label{item:lessk1} If $i < k-1$, $B$ outputs $p_i =
  \Adv^*(1^n,r_1,r_2,\ldots,r_{i};z)$ (i.e., $B$ proceeds just as
  $\Adv^*$ in the first $2k-4$ rounds).
\item\label{item:equalk1} If $i = k-1$ (i.e., in round $2k-2$), $B$
  lets $(p_{k-1},p_{k}^1,\ldots,p_{k}^{\reps}) =
  \Adv^*(1^n,r_1,r_2,\ldots,r_{k-1},\widr_{k}^1,\ldots,\widr_{k}^{\reps};z)$
  and outputs $p_{k-1}$.
\item If $i = k$ (i.e., in round $2k$), then:
\BI

\item $B$ lets $T = (r_1,p_1,\ldots,p_{k-2},r_{k-1},
  \widr^1_{k},\ldots,\widr^\reps_{k},p_{k-1},p^1_k,\ldots,p^\reps_k)$,
  and lets $y =  (pad,T_{\leq  k-1},p_{k-1},r_k)$ if $T$ is accepting and $y =
    \bot$ otherwise.

\item $B$ lets $u \leftarrow \Inv(y)$ and interprets $u$ as $(pad,\widi,r_M)$ where
$|pad|=n$, $|\widi| = \log_2(\reps)$, and
$r_M=(z',\hatr_1,\hatr_2,\ldots,\hatr_{k-1},\hatr_{k}^1,\ldots,\hatr_{k}^{\reps})$,
such that $z' \in \{0,1\}^{\rta}$ and $\hatr_1,\hatr_2,\ldots,\hatr_{k-1},\hatr_{k}^1,\ldots,\hatr_{k}^\reps$ are in $\{0,1\}^{\ell_c}$. 
\item
$B$ next lets
$(\widp_{k-1},\widp_{k}^1,\ldots,\widp_{k}^{\reps})  = 
  \Adv^*(1^n,r_1,r_2,\ldots,r_{k-1},\hatr_{k}^1,\ldots,\hatr_{k}^{\reps};z')$
  and outputs $\widp^\widi_{k}$.
\EI
\EE
We now proceed to analyze the success probability of $B$ against $\Chal$.  
In particular, we shall show that 
for all sufficiently large $n$, $\defsuc{B}{n} >\frac{1}{64}$
which will conclude the proof of Lemma \ref{lem:collapse}.
We denote by $\view_{\Adv^*}(\inter{\Adv^*,\nv}(1^n))$ the random
variable that represents the view of the adversary $\Adv^*$ in an
interaction with $\nv$ on common input $1^n$---for convenience, we
describe this view $v = (z,T)$ by $\Adv^*$'s random coin tosses $z$, as well as the
transcript $T$ of the interaction between $\Adv^*$ and
$\nv$.\footnote{This is a bit redundant---as the messages sent by
  $\Adv^*$ can of course be recomputed given just the randomness of
  $\Adv^*$ and the messages from $\nv$, but will simplify notation.}
Towards analyzing $B$, we consider a sequence of hybrid experiments
$\Expt_0,\Expt_1,\Expt_2,\Expt_3$---formally, an experiment defines a
probability space and a probability density function over it. All
experiments will be defined over the same probability space so we can
consider the same random variables over all of them. To simplify
notation, we abuse of notation and let
$\Expt_i(n)$ also denote a random variable describing the output of the
experiment $\Expt_i(n)$.

$\Expt_0(n)$ will simply consider an
execution between $B^{\Inv}$ and $\Chal$ on common input $1^n$ and will
output  1 if $\Chal$ is accepting and 0 otherwise; see
Figure~\ref{fig:experiments} for a formalization. To simplify the
transition to later experiments, we formalize $\Expt_0(n)$ as first
sampling a full transcript $T$ of an execution between $\Adv^*$ and
$\nv$, keeping only the prefix $T_{\leq k-1}$ (this gives exactly
the same distribution as an interaction between $B$ and $\Chal$ up to
round $2k-2$), next
sampling an ``external'' random message $r$ (just as $\Chal$ would
in round $2k-1$), and finally producing the last message just as $B$
does. (We additionally sample a random index $i \in [\reps]$ is not
used in the current experiment, but will be useful in later experiments.)
We thus directly have:
\BCM\label{clm:hybreal} $\defsuc{B}{n} = \Pr[\Expt_0(n)=1]$
\ECM
We now slowly transform the experiment into one that becomes easy to
analyze. See Figure~\ref{fig:experiments} for a
formal description of the experiments.
\begin{figure}
\begin{boxfig}{H}{
\textbf{Experiment $\Expt_0(n).$}
\begin{enumerate}[\hspace{0.05cm}1.]
\item Sample $(z,T) \leftarrow
  \view_{\Adv^*}(\inter{\Adv^*,\nv}(1^n)); pad\leftarrow \{0,1\}^n; i
  \leftarrow[\reps]; {r \leftarrow \{0,1\}^{\ell_c}}$.
  Interpret $T$ as $(r_1,p_1,\ldots,p_{k-2},r_{k-1}, \widr^1_{k},\ldots,\widr^\reps_{k},p_{k-1},p^1_k,\ldots,p^\reps_k)$. 
\item Let $r_k = r$ and let {$y =
    (pad,T_{\leq  k-1},p_{k-1},r_k)$ if $T$ is accepting and $y =
    \bot$ otherwise}.
  \item Let $u \leftarrow \Inv(y)$; interpret $u$ as $(pad,\widi,r_M)$ where $|pad|=n$, $|\widi| =
    \log_2(\reps)$, and
    $r_M=(z',\hatr_1,\hatr_2,\ldots,\hatr_{k-1},\hatr_{k}^1,\ldots,\hatr_{k}^{\reps})$
    just as $B$ does and let 
    $(\widp_{k-1},\widp_{k}^1,\ldots,\widp_{k}^{\reps}) =
    \Adv^*(1^n,r_1,r_2,\ldots,r_{k-1},\hatr_{k}^1,\ldots,\hatr_{k}^{\reps};z')$.
    \item Output 1 iff $T' =  (T_{\leq k-1}, p_{k-1}, r_k, \widp_{k}^\widi)$
      is accepting, and 0 otherwise.
\end{enumerate}
	
\hrulefill	\\
\textbf{Experiment $\expt_1(n).$}
\begin{enumerate}[\hspace{0.05cm}1.]
\item Sample $(z,T) \leftarrow
  \view_{\Adv^*}(\inter{\Adv^*,\nv}(1^n)); pad\leftarrow \{0,1\}^n; i
  \leftarrow[\reps]; {r \leftarrow \{0,1\}^{\ell_c}}$.
  Interpret $T$ as $(r_1,p_1,\ldots,p_{k-2},r_{k-1}, \widr^1_{k},\ldots,\widr^\reps_{k},p_{k-1},p^1_k,\ldots,p^\reps_k)$. 
\item Let \textcolor{blue}{$r_k = r$} and let {$y =
    (pad,T_{\leq  k-1},p_{k-1},r_k)$ if $T$ is accepting and $y =
    \bot$ otherwise}.
  \item Let $u \leftarrow \Inv(y)$; interpret $u$ as $(pad,\widi,r_M)$ where $|pad|=n$, $|\widi| =
    \log_2(\reps)$,
and
    $r_M=(z',\hatr_1,\hatr_2,\ldots,\hatr_{k-1},\hatr_{k}^1,\ldots,\hatr_{k}^{\reps})$
    just as $B$ does and let 
    $(\widp_{k-1},\widp_{k}^1,\ldots,\widp_{k}^{\reps}) =
    \Adv^*(1^n,r_1,r_2,\ldots,r_{k-1},\hatr_{k}^1,\ldots,\hatr_{k}^{\reps};z')$.
    \item Output 1 iff $T' =  (T_{\leq k-1}, p_{k-1}, r_k, \widp_{k}^\widi)$
      is accepting \textcolor{orange}{and $G$ holds}, and 0 otherwise.
\end{enumerate}

\hrulefill	\\
\textbf{Distribution $ \expt^n_2$}
\begin{enumerate}[\hspace{0.05cm}1.]
\item Sample $(z,T) \leftarrow
  \view_{\Adv^*}(\inter{\Adv^*,\nv}(1^n)); pad\leftarrow \{0,1\}^n; i
  \leftarrow[\reps]; {r \leftarrow \{0,1\}^{\ell_c}}$.
  Interpret $T$ as $(r_1,p_1,\ldots,p_{k-2},r_{k-1}, \widr^1_{k},\ldots,\widr^\reps_{k},p_{k-1},p^1_k,\ldots,p^\reps_k)$. 
\item Let \textcolor{blue}{$r_k = \widr^i_k$} and let {$y =
    (pad,T_{\leq  k-1},p_{k-1},r_k)$ if $T$ is accepting and $y =
    \bot$ otherwise}.
  \item Let $u \leftarrow \textcolor{red}{\Inv(y)}$; interpret $u$ as $(pad,\widi,r_M)$ where $|pad|=n$, $|\widi| =
    \log_2(\reps)$, 
and
    $r_M=(z',\hatr_1,\hatr_2,\ldots,\hatr_{k-1},\hatr_{k}^1,\ldots,\hatr_{k}^{\reps})$
    just as $B$ does and let 
    $(\widp_{k-1},\widp_{k}^1,\ldots,\widp_{k}^{\reps}) =
    \Adv^*(1^n,r_1,r_2,\ldots,r_{k-1},\hatr_{k}^1,\ldots,\hatr_{k}^{\reps};z')$.
    \item Output 1 iff $T' =  (T_{\leq k-1}, p_{k-1}, r_k, \widp_{k}^\widi)$
      is accepting and $G$ holds, and 0 otherwise.
\end{enumerate}

 \hrulefill\\
 \textbf{Distribution $\expt^n_3$}

\begin{enumerate}[\hspace{0.05cm}1.]
\item Sample $(z,T) \leftarrow
  \view_{\Adv^*}(\inter{\Adv^*,\nv}(1^n)); pad\leftarrow \{0,1\}^n; i
  \leftarrow[\reps]; {r \leftarrow \{0,1\}^{\ell_c}}$.
  Interpret $T$ as $(r_1,p_1,\ldots,p_{k-2},r_{k-1}, \widr^1_{k},\ldots,\widr^\reps_{k},p_{k-1},p^1_k,\ldots,p^\reps_k)$. 
\item Let {$r_k = \widr^i_k$} and let {$y =
    (pad,T_{\leq  k-1},p_{k-1},r_k)$ if $T$ is accepting and $y =
    \bot$ otherwise}.
  \item Let $u \leftarrow \textcolor{red}{\PInv(y)}$; interpret $u$ as $(pad,\widi,r_M)$ where $|pad|=n$, $|\widi| =
    \log_2(\reps)$,
and
    $r_M=(z',\hatr_1,\hatr_2,\ldots,\hatr_{k-1},\hatr_{k}^1,\ldots,\hatr_{k}^{\reps})$
    just as $B$ does and let 
    $(\widp_{k-1},\widp_{k}^1,\ldots,\widp_{k}^{\reps}) =
    \Adv^*(1^n,r_1,r_2,\ldots,r_{k-1},\hatr_{k}^1,\ldots,\hatr_{k}^{\reps};z')$.
    \item Output 1 iff $T' = (T_{\leq k-1}, p_{k-1}, r_k, \widp_{k}^\widi)$
      is accepting and $G$ holds, and 0 otherwise.
\end{enumerate}
}
 \end{boxfig}
\caption{Description of intermediate experiments.}\label{fig:experiments}
\end{figure}

\BE
\item We first define an a ``good'' event $G = W \cap G'$, where $W$
  is the event that the originally sampled transcript is accepting and
  $G'$ is the event that the ``prefix'' $(T_{\leq
    k-1, p_{k-1}})$ is ``good''  in a well defined sense
  (roughly speaking, that continuations conditioned $T_{\leq
    k-1}$ are successful with high probability, and that that in such
  successful continuations $p_{k-1}$ is used with not ``too low''
  probability).
  $\Expt_1$ will next proceeds
  just like $\Expt_0$ except that we additionally fail if the event $G$ does not
  happen.
  We thus have that the probability of $\expt_0(n)$ outputting 1 is at
  least as high as the probability of  $\expt_1(n)$ outputting 1.
  \BCM\label{clm:hyb1}
  $\Pr[\Expt_0(n)=1] \geq \Pr[\Expt_1(n)=1].$
\ECM
  Additionally, as we shall show (using an averaging argument), the
  event $G$ happens with non-negligible probability, not just in
  $\Expt_1$ but also in all the other experiments $\Expt_j$ for $j \in
  \{1,2,3\}$ (as Step 1 of the experiment remains unchanged in all of them).
  \BCM\label{clm:bndg} For $j \in \{1,2,3\}$, 
  $\Pr[G] \geq \frac{\epsilon^2}{8}$, 
where the probability is over the randomness in experiment $\Expt_j(n)$
\ECM

  \item We next transition to an experiment $\Expt_2$ where instead of
    choosing the message $r_k$ at random (as it was in $\Expt_1$), we
    select it as the message in the $i$'th repetition of $\nv$'s
    $k-1$'st message in the initially sampled
    transcript $T$, where $i$ is a randomly sampled index $i \in [\reps]$. The reason for defining this experiment is that, in it,
    we are applying the one-way function inverter on the ``right''
    distribution (just as in the definition of $M$).
    The central claim to show is that this change does not change the
    success probability by too much. As discussed in the introduction,
    we shall prove it using Raz's sampling lemma. 
    \BCM\label{clm:hyb2}
 $\Pr[\Expt_1(n)=1] \geq  \Pr[\Expt_2(n)=1] - \frac{1}{\log(n)}.$
\ECM
\item Finally, we transition to an experiment $\Expt_2$ where we employ
  a \emph{perfect inverter} $\PInv$---that always samples uniform preimages
  to $M$, instead of the (imperfect) inverter $\Inv$. It directly
  follows from the fact that $\Inv$ is an inverter with statistical
  closeness $\frac{1}{n}$ and that the inverter is applied to an element that
  is sampled as a uniform image of $M$\footnote{Note that we here rely
    on the fact that $y = \bot$ when $T$ is not accepting.} that the statistical distance
  between $\Expt_2(n)$ and $\Expt_3(n)$ is bounded by $\frac{1}{n}$ for
  sufficiently large $n$. In particular,
\BCM\label{clm:hyb3}
For all sufficiently large $n$, 
$$\Pr[\Expt_2(n)=1] \geq \Pr[\Expt_3(n)=1] - \frac{1}{n}$$
\ECM
\item We finally note that in $\Expt_3$, there are only two reason the
  experiment can output 0: (1) The originally sampled transcript $T$
  is not accepting (i.e., the event $W$ does not hold); if is is
  accepting, the perfect inverter will make sure that $T'$ is also accepting, or (2) the
  event $G$ does not hold.
  Additionally note, since $G$ is defined as $W \cap G'$, we have that
  whenever $G$ holds, $W$ holds as well and thus the experiment must
  output 1. Thus, by Claim \ref{clm:bndg}, we have:
\BCM\label{clm:hyb4}
$$\Pr[\Expt_3(n)=1] \geq \frac{\eps^2}{8}$$
\ECM
\item By combining claims \ref{clm:hybreal}, \ref{clm:hyb1},
  \ref{clm:hyb2}, \ref{clm:hyb4},
  we have that for all sufficiently large $n$, 
  \begin{align*}
    \defsuc{B}{n} &=\Pr[\Expt_0(n)=1] \geq \Pr[\Expt_1(n)=1] \geq
  \Pr[\Expt_2(n)=1] - \frac{1}{\log(n)} \\ &\geq \Pr[\Expt_3(n)=1] -
                                             \frac{2}{\log(n)}
    \geq  \frac{\eps^2}{8} - \frac{2}{\log(n)} > \frac{1}{64}
    \end{align*}
which is a contradiction.
\EE
To to conclude the proof of Lemma \ref{lem:collapse}, it just remains
to formalize the event $G = W \cap G'$ and proving
Claim \ref{clm:bndg} and Claim \ref{clm:hyb3}.

\subsection{The Good Event $G$ and the Proof of Claim \ref{clm:bndg}}\label{sec:defineg}
We begin by defining some random
variables over the probability space over which $\Expt_1$ is defined.
Note that the probability space is the same for
$\Expt_0,\Expt_1,\Expt_2$ and as such random variables and events over
$\Expt_1$ are also defined over all the other experiments.
We use boldface to denote random variables describing the outcome of
variables in the experiments---for instance, we let ${\bf T}$ denote a random variable
describing the value of $T$ as sampled in the experiments. 

Let $W$ denote the event that ${\bf T}$ is accepting (i.e., the
transcript sampled in Step 1 is accepting) and let $\Theta$ be the set of partial transcripts $\theta$ such that 
$$\Pr [W \mid {\bf T}_{\leq(k-1)} = \theta] \geq
\frac{\eps}{2}.$$ where the probability is over $\Expt_1(n)$. 
That is, $\Theta$ is the set of ``good'' partial transcripts conditioned on
which $\Adv^*$ has a reasonable probability of succeeding.
Note that by a standard averaging argument, we have that such
transcripts occur often:
\begin{equation}
\Pr[{\bf T}_{\leq(k-1)} \in \Theta] \geq \frac{\eps}{2}. \label{eqn:bndtheta}
\end{equation}
Now, consider the event $W_p$ that $W$ holds and ${\bf p}_{k-1} = p$; let
$P(\theta)$ be the set of messages $p \in \{0,1\}^{\ell_a}$ for which
$$\Pr[W_p \mid {\bf T}_{\leq(k-1)}=\theta] \geq \frac{\eps}{2^{\ell_a+2}}.$$
In other words, $P(\theta)$ is the set of ``good'' (adversary) messages $p$
such that conditioned on the partial transcript $\theta$, the probability that $\Adv^*$ succeeds
while using $p$ as its $k-1$'st message is greater than $\frac{\eps}{2^{\ell_a+2}}$. 
As we shall now show using another (standard) averaging argument, for
every $\theta \in \Theta$, we have
\begin{equation}
\Pr[ {\bf p}_{k-1} \in P(\theta) \mid {\bf T}_{\leq(k-1)}=\theta] \geq \frac{\eps}{4}. \label{eqn:bndp}
\end{equation}
Suppose for contradiction that for some $\theta \in \Theta$, Equation~\ref{eqn:bndp} does not hold. Then, we have
\begin{align*}
  \Pr[W \mid {\bf T}_{\leq(k-1)}=\theta]&= 
  \sum_{p \in \{0,1\}^{\ell_a}} \Pr[W_p \mid {\bf
  T}_{\leq(k-1)}=\theta] \\
&= \sum_{p \in P(\theta)}\Pr[W_p\mid {\bf T}_{\leq(k-1)}=\theta ]
+ \sum_{p \in \{0,1\}^{\ell_a}-P(\theta)}\Pr[W_p \mid {\bf
  T}_{\leq(k-1)}=\theta ] \\
& \leq \Pr[{\bf p}_{k-1} \in P(\theta)\mid {\bf T}_{\leq(k-1)}=\theta ]
+ \sum_{p \in \{0,1\}^{\ell_a}-P(\theta)}\Pr[W_p \mid {\bf
  T}_{\leq(k-1)}=\theta ]   \\
&< \frac{\epsilon}{4}
  + \sum_{p \in \{0,1\}^{\ell_a}-P(\theta)}\Pr[W_p \mid {\bf
  T}_{\leq(k-1)}=\theta ]    \\
&\leq \frac{\eps}{4}  +\sum_{p \in \{0,1\}^{\ell_a}-P(\theta)}
  \frac{\eps}{2^{\ell_a+2}} \\
&\leq \frac{\eps}{4} +2^{\ell_a}\cdot  \frac{\eps}{2^{\ell_a+2}} = \frac{\eps}{2}
\end{align*}
which contradicts that $\theta\in\Theta$. 

Next, define $G'$ to be the event that ${\bf T}_{\leq(k-1)} \in
\Theta$ and ${\bf p}_{k-1} \in P({\bf T}_{\leq(k-1)})$, and define $G$
as holding when $W$ and $G'$ both hold (i.e., the originally sampled transcript is accepting
and $G'$ holds). Note that $G'$ in fact implies that $W$ holds (since ${\bf p}_{k-1} \in P({\bf
  T}_{\leq(k-1)})$ implies that ${\bf p}_{k-1} \neq \bot$ which by our
assumption on $A^*$ means that ${\bf T}$ must be accepting), thus in
fact $G' = G$.
By combing Equations \ref{eqn:bndtheta} and \ref{eqn:bndp}, we have:
\begin{align*}
  \Pr[G] = \Pr[G'] &=\Pr[{\bf T}_{\leq(k-1)} \in
\Theta \land {\bf p}_{k-1} \in P({\bf T}_{\leq(k-1)})]
\\&=
  \Pr[{\bf T}_{\leq(k-1)} \in
\Theta] \times \Pr[{\bf p}_{k-1} \in P({\bf T}_{\leq(k-1)}) \mid {\bf T}_{\leq(k-1)} \in
\Theta] \\&\geq \frac{\epsilon}{2} \times
          \frac{\epsilon}{4}  = \frac{\eps^2}{8}
\end{align*}
where the probability is taken over $\Expt_1(n)$. 
Finally, note that
since Step 1 (whose outcome determines whether $G$ happens) remains unchanged in all the experiments, we can
conclude that $\Pr[G] \geq \frac{\eps^2}{8}$ where the probability is
taken over $\Expt_j(n)$ for every $j \in \{1,2,3\}$,  which concludes the proof of Claim~\ref{clm:bndg}.

\subsection{Proof of Claim \ref{clm:hyb2}}
Recall that we need to show that $\Pr[\Expt_1(n)=1] \geq
\Pr[\Expt_2(n)=1] - \frac{1}{\log(n)}$. Observe that the only difference between experiments $\Expt_1$ and
$\Expt_2$ is that, in  $\Expt_1$, we set $r_k = r$ and in $\Expt_2$,
we set $r_k = \widr_k^i$.  Furthermore,  both the experiments sample
$((z,T),pad,i,r)$ from the same distributions and output $0$ whenever
$G$ does not hold (which is a function only of $T$). It follows that the statistical distance between
$\Expt_1(n)$ and $\Expt_2(n)$ is bounded by the statistical distance
of $\Expt_1(n)$ and $\Expt_2(n)$ conditioned on the event $G$. 
Note that we can rephrase the event $G$ as 
$$G = \bigcup_{\theta \in \Theta, p \in P(\theta)} W_p \cap ({\bf
  T}_{\leq k-1} = \theta)$$ 
Below, we shall
show that for every $\theta \in \Theta, p \in P(\theta)$, it holds that
the statistical distance between $\{\Expt_1(n) \mid {\bf T}_{\leq k-1} = \theta,
    W_p\}$ and $\{\Expt_2(n) | {\bf T}_{\leq k-1} = \theta, W_p\}$ is bounded by $\frac{1}{\log(n)}$, which concludes
    the proof of Claim \ref{clm:hyb3}.

    Consider some $\theta \in \Theta, p \in P(\theta)$
and 
consider
    the experiments $\{\Expt_1(n) \mid {\bf T}_{\leq k-1} = \theta,
    W_p\}$ and $\{\Expt_2(n) \mid {\bf T}_{\leq k-1} = \theta, W_p\}$. Note both experiments proceed exactly the
    same after $r_k$ is defined in Step 2, so we can ignore everything that happens after
    this. Additionally, note that the only variables that are
    relevant after this point are ${\bf pad}, {\bf T}_{\leq k-1},
    {\bf p}_{k-1}$, ${\bf i}$ and ${\bf r}$. Note that ${\bf pad},
    {\bf i}$ are both independent of the events ${\bf T}_{\leq k-1} = \theta,W_p$ and thus
    still independently and uniformly sampled in both
    experiments. ${\bf T}_{\leq k-1}$ and ${\bf p}_{k-1}$, on the
    other hand are fixed (constant)
    conditioned on ${\bf T}_{\leq k-1} = \theta,W_p$. Thus, to bound the statistical
    difference between $\{\Expt_1(n) \mid {\bf T}_{\leq k-1} = \theta, W_p\}$ and $\{\Expt_1(n)
        \mid {\bf T}_{\leq k-1} = \theta,W_p\}$, it suffices to bound the statistical distance
        between ${\bf r_k}$ in $\{\Expt_1(n)\mid {\bf T}_{\leq k-1} =
        \theta,W_p\}$ and ${\bf r_k}$ in $\{\Expt_2(n) \mid {\bf T}_{\leq k-1} = \theta,W_p\}$.
        In other words, we need to upper bound,
        $$\Delta = \mathsf{SD}({\bf r}, {\bf s}^{\bf i}_k \mid  W_p) =
        \mathsf{SD}({\bf s}^{\bf i}_k, {\bf s}^{\bf i}_k \mid  W_p)
        \leq \sum_{j\in [\reps]}  \frac{1}{\reps} \mathsf{SD}({\bf s}^j_k, {\bf s}^j_k \mid  W_p)
 $$
        over $\{\Expt_2(n) \mid {\bf T}_{\leq k-1} = \theta\}$ since
        for each $j$, ${\bf s}^j_k$ is sampled uniformly at random,
        independent of $T_{\leq k-1}$ (and independent of ${\bf
          s}^{j'}_k$ for $j' \neq j$).
        Towards bounding this quantity, we will rely on Raz's sampling lemma.

\BL[\cite{Raz98}]\label{lem:raz} 
Let ${\bf X}_1,\ldots,{\bf X}_,$ be independent random variables on a
finite domain $U$. Let $E$ be an event over $\vec{{\bf X}} = ({\bf
  X}_1,\ldots,{\bf X}_m)$. Then, 
$$\frac{1}{m}\cdot \sum_{i=1}^m \mathsf{SD}({\bf X}_i,{\bf X}_i|E) \leq \sqrt{\frac{1}{m}\cdot \log\frac{1}{\Pr[E]}}$$
\EL
\noindent By applying Raz's lemma, we directly get that
$$\Delta \leq
\sqrt{\frac{1}{\reps}\cdot \log\frac{1}{\Pr[W_p]}}$$ where the
probability is over $\{\Expt_2(n) \mid {\bf T}_{\leq k-1} = \theta\}$. 
Since by our assumption $p \in P(\theta)$, we have that the
probability of $W_p$ conditioned on ${\bf T}_{\leq k-1} = \theta$
is at least $\frac{\epsilon}{2^{\ell_a+2}}$, thus
$$\Delta  \leq \sqrt{\frac{1}{\reps}\cdot (\ell_a+2-\log(\eps))} \leq \frac{1}{\log(n)}$$
since  $\eps > \frac{1}{2}$ and $\reps= (\ell_a+4)(\log(n))^2> (\ell_a+2-\log(\eps))(\log(n))^2$.

\EPF

\subsection{Variations}
Using essentially the same proofs, we can directly get the following
vacations of \ref{lem:collapse}. The first variant simply states that
the same result holds for almost-everywhere puzzles.
\BL [Almost-everywhere variant 1] \label{lem:collapse:var1}
Assume there exists a $k(\cdot)$-round almost-everywhere
public-coin puzzle such that $k(n)\geq 3$.
Then, either there exists an ioOWF, or there exists a
$(k(\cdot)-1)$-round almost-everywhere 
public-coin puzzle. Moreover, if the $k(\cdot)$-round puzzle has perfect
completeness, then either there exists an ioOWF, or a $(k(\cdot)-1)$-round almost-everywhere public-coin puzzle with perfect-completeness. 
\EL
The next variant shows that if we start off with an almost-everywhere
puzzle, we can either get a (standard) one-way function or a puzzle
with one less round (but this new puzzle no longer satisfies
almost-everywhere security)
i
This follows from the
fact that if the attacker $A^*$ succeeds on all sufficiently large
input lengths, then it suffices for $\Inv$ to work on infinitely many
input lengths, to conclude that $B^{\Inv}$ works on infinitely many
inputs length (thus violating almost-everywhere security of the
original puzzle).
\BL [Almost-everywhere variant 2] \label{lem:collapse:var2} 
Assume there exists a $k(\cdot)$-round almost-everywhere
public-coin puzzle such that $k(n)\geq 3$.
Then, either there exists a OWF, or there exists a
$(k(\cdot)-1)$-round public-coin puzzle. Moreover, if the $k(\cdot)$-round puzzle has perfect
completeness, then either there exists a OWF, or a $(k(\cdot)-1)$-round public-coin puzzle with perfect-completeness. 
\EL
We additionally consider a variant for non-uniform puzzles. As the
challenger now may be a non-uniform $\PPT$, the function $M$ that we
are required to invert is also a non-uniform $\PPT$ and thus we can
only conclude the existence of non-uniform OWFs.
\BL [Non-uniform variant] \label{lem:collapse:var3}
Assume there exists a $k(\cdot)$-round non-uniform public-coin puzzle such that $k(n)\geq 3$.
Then, either there exists a non-uniform ioOWF, or there exists a
$(k(\cdot)-1)$-round non-uniform 
public-coin puzzle.\footnote{The transformation still preserves perfect
  completeness, but this will not be of relevance for us.} 
\EL

\subsection{Characterizing O(1)-Round Public-coin Puzzles}
We next apply our round-collapse theorem (and its variants) to get a
characterization of $O(1)$-round puzzles. This characterization
applies to both standard puzzles and non-uniform puzzles.
\BCR\label{cor:kto2}
Assume the existence of a $O(1)$-round (resp. a $O(1)$-round
non-uniform) public-coin puzzle. Then there exists a
$2$-round public-coin puzzle (resp. $2$-round non-uniform public-coin puzzle) and thus a distributional $\NP$ problem (resp. distributional $\NPpoly$ problem)
 that is HOA (resp. nuHOA).
\ECR
\BPF
If (non-uniform) ioOWF exists, then by applying Proposition \ref{lem:2puzowf} we have that
2-round (non-uniform) public-coin puzzles exist.
If (non-uniform) ioOWF do not exist, we can apply
Lemma~\ref{lem:collapse} (Lemma \ref{lem:collapse:var3}) iteratively to collapse any
constant-round protocol to a 2-round protocol. (Note that we can only
apply Lemma~\ref{lem:collapse} a constant number of times, as the
communication complexity of the resulting protocol grows polynomially
with each application.). Thus in either case, we conclude that the
existence of a $O(1)$-round (non-uniform) public-coin puzzle implies a 2-round
(non-uniform) public-coin puzzle. The corollary is concluded by applying
Lemma \ref{lem:puz2hoa}. 
\EPF

We remark that the reason we cannot get an (unconditional)
characterization of almost-everywhere
puzzles is that ioOWFs.
are not known to imply 2-round almost-everywhere puzzles.

\section{Characterizing Polynomial-round Puzzles}
We observe that the existence of a $\poly$-round public-coin puzzle is
equivalent to the statement that $\PSPACE \not\subseteq \BPP$. A consequence of this
result (combined with Lemma~\ref{lem:puz2hoa}) is that any round-collapse theorem that (unconditionally)
can transform a polynomial-round puzzle into a $O(1)$-round puzzle,
must show the existence of a HAO distributional $\NP$ problem based on
the assumption that $\PSPACE \not\subseteq \BPP$ (which would be
highly unexpected).

\newcommand{\rta}{h}
\BT\label{thm:pspace}
For every $\epsilon>0$, there exists an $n^\epsilon$-round public-coin puzzle
(resp. a non-uniform puzzle) if and only if $\PSPACE
\not\subseteq \BPP$ (resp. $\PSPACE
\not\subseteq \Ppoly$).
\ET
\BPF
For the ``only-if'' direction, note that using the same proof as (the
easy direction) in $\IP = \PSPACE$
\cite{Shamir92,LundFKN92}, we can use a $\PSPACE$ oracle to implement
the optimal adversary strategy in every puzzle and thus (due to the
completeness condition of the puzzle) break the soundness of every puzzle using a
$\PSPACE$ oracle. So, if $\PSPACE \subseteq \BPP$, soundness of every
puzzle can be broken in $\PPT$ and thus puzzles cannot exist. (We
remark that a very similar statement---in the language of non-trivial interactive
arguments---was already observed by Goldreich \cite{oded18}; see
Section \ref{nontrivial.sec} for more details.)

For the ``if'' direction, recall that by the classic result of \cite{BabaiFNW93} (see also
\cite{TrevisanV07}), if $\PSPACE \not\subseteq \BPP$, then there is
$\PSPACE$ language $L'$, constant $c \in \NN$, and a polynomial $p(\cdot)$   
such that $(L',\cU_p)$  is $\frac{1}{n^c}$-HOA. We will now use this
HAO language $L'$ together with the fact that by
\cite{Shamir92,LundFKN92} all of $\PSPACE$ has a public-coin
interactive proof (and the fact that $\PSPACE$ is closed under
complement, 
to get a puzzle. The puzzle challenger $\Chal(1^n)$ simply samples a random
statement $x \in \{0,1\}^{p(n)}$ and sends it to the adversary. The
adversary next announces a bit $b$ (determining whether $x \in L'$ or
not) and next if $b =1$, $\Chal$ runs the IP verifier for $x\in L'$ and
if $b=0$ instead runs the IP verifier for $x \notin L'$. Due to
\cite{Shamir92,LundFKN92}, we may assume without loss of generality
that the IP has completeness 1 and soundness error $2^{-n}$. As we
shall now argue $\Chal$ is a $(1,1-\frac{2}{n^c})$-puzzle which by
remark Remark~\ref{rem:par} implies a puzzle. Completeness follows
directly from the completeness of the IP. For soundness, consider a
$\ppt$ machine $\Adv^*$ that convinces $\Chal$ with probability better
than $1-\frac{2}{n^c}$. We construct a machine $B$ that breaks the
$\frac{1}{n^c}$-HAO property of $L'$. $B(1^n,x)$ simply emulates an
interaction between $\Chal(1^n)$ and $\Adv^*$ while fixing $\Chal$'s
first message to $x$ and accepts $x$ if $\Chal$ is accepting, and
rejects otherwise. Since $B$ is feeding $\Adv^*$ messages according to
the same distribution as in the real execution (with $\Chal$), we have
that $\Adv^*$ convinces $\Chal$ in the emulation by $B$ with
probability at least $1-\frac{2}{n^c}$. By the soundness of the IP, we
have that except with probability $2^{-n}$, whenever the proof
is accepting, the bit $b$ must correctly decide $x$. We conclude (by a
union bound) that $B$ correctly decides $x$ with probability
$1-\frac{2}{n^c} - 2^{-n} > 1-\frac{1}{n^c}$  for all sufficiently
large $n \in \NN$.

The non-uniform version of the theorem follows using exactly the same
proof.
\EPF

\section{Achieving Perfect Completeness}
We show that any 2-round public-coin puzzle can be transformed into a 3-round
public-coin puzzle with perfect completeness; next, we shall use this result
together with our round-reduction theorem to conclude our main result.

\subsection{From Imperfect to Perfect Completeness (by Adding a Round)}
Furer et al. \cite{FurerGMSZ89} showed how to transform any 2-round
public-coin proof system into a 3-round public-coin proof system with
perfect completeness. We will rely on the same protocol transformation to
transform any $2$-round puzzle into a $3$-round puzzle with perfect
completeness. The perfect completeness condition will follow directly
from their proof; we simply must argue that the transformation also
preserves \emph{computational} soundness (as they only showed that it
preserves information-theoretic soundness).

\BT\label{thm:2toperfect}
Suppose there exists $2$-round public-coin puzzle. Then there exists a $3$-round public-coin puzzle with perfect completeness. 
\ET
\BPF  Let $\myprot{\Adv}{\Chal}$ be a 2-round public-coin puzzle.  Let
$\ell_c,\ell_a$ be polynomials such that the
message from $\Chal(1^n)$ is of length $\ell_c(n)$ and the
message from $\Adv(1^n)$ is of length $\ell_a(n)$; we assume without
loss of generality that $\ell_c(n)>2$. When the security parameter $n$
is clear from the context we will omit it and let $\ell_c(n) = \ell_c$
and $\ell_a(n) = \ell_a$.

We now apply the Furer et al. \cite{FurerGMSZ89} transformation to this
puzzle to create a $3$-round puzzle $\myprot{\np}{\nv}$. The puzzle
will proceed by first having the adversary sending $\ell_c$ ``pads''
$z_1,\ldots,z_{\ell_c} \in {\ell_c}$ to $\nv$; $\nv$ next sends back a
random message $r_{\nv} \in \{0,1\}^{\ell_c}$, and the adversary is
next supposed to find a response $i,p$ such that $(r \oplus z_i, p)$
is a valid transcript for the original puzzle (i.e., the adversary
needs to win in one of the parallel ``padded'' instances of the
original puzzle). More formally,
$\nv(1^n,(z_1,\ldots,z_{\ell_c}),(i,p);r_{\nv}) = 1$ if and only if $\Chal(1^n,p;r_{\nv}\oplus z_i)$ outputs 1.
Perfect completeness of $\nv$ follows directly from the original proof
by \cite{FurerGMSZ89}. For completeness, we recall their proof.
From the (imperfect) completeness of $\Chal$, we have that there
exists some adversary $\Adv$ such that 
 $\defsuc{\Adv}{n} \geq 1 - \frac{1}{n}$ for all sufficiently
 large $n$; without loss of $A$ is deterministic. Fix some $n>2$ for
 which this holds. Let $S\subseteq
 \{0,1\}^{\ell_c}$ be the set of challenges for which $\Adv$ provides an
 accepting response; the probability that a random challenge $z \in
 \{0,1\}^{\ell_c}$ is inside $S$
 is thus at least $1 - \frac{1}{n}$.  
 We
 will show that there exists ``pads'' $z_1,\ldots,z_{\ell_c}$ such
 that for every $r \in \{0,1\}^{\ell_c}$, there exists some $i$ such that $r
 \oplus z_i \in S$, which concludes that an unbounded attacker $\np$ can
 succeed with probability 1 (by selecting those pads and next
 providing an accepting response).
Note that for every fixed $r$, for a \emph{randomly} chosen pad $z_i$, the
probability that $r \oplus z_i \notin S$ is at most
$\frac{1}{n}$; and thus the probability over randomly chosen
pads $z_1, \ldots, z_{\ell_c}$ that $r \oplus z_i \notin S$  \emph{for all}
$i$ is at most $\frac{1}{n^{\ell_c}}$. We conclude, by a union bound,
that the probability over randomly chosen
pads $z_1, \ldots, z_{\ell_c}$ that \emph{there exists} some $r \in
\{0,1\}^{\ell_c}$ such that $r \oplus z_i \notin S$ for all
$i$ is at most $\frac{2^{\ell_c}}{n^{\ell_c}} < 1$.  Thus, there
exists pads $z_1, \ldots, z_{\ell_c}$ such that \emph{for every } $r \in
\{0,1\}^{\ell_c}$ there exists some $i$ such that $r \oplus z_i \notin
S$, which concludes perfect completeness.

We now turn to proving computational soundness.
Consider some adversary $\np^*$
that succeeds in convincing $\nv$ with probability $\epsilon(n)$ for
all $n \in \NN$. We construct an adversary $\Adv^*$ that convinces $\Chal$ with
probability $\frac{\epsilon(n)}{\ell_c}$, which is a
contradiction. $\Adv^*(1^n)$ picks a random tape $r_{\np^*}$ for
$\np^*$, lets $(z_1,\ldots,z_{\ell_c}) = \np^*(1^n;r_{\np^*})$, picks
a random index $i \in [\ell_c]$ and outputs $z_i$. Upon receiving
 a ``challenge'' $r$, it lets $(j,p) = \np^*(1^n,r\oplus z_i;r_{\np^*})$
outputs $p$ if $i = j$ and $\bot$ otherwise.
First, note that in the emulation by $\Adv^*$, $\Adv^*$ feeds $\np^*$
the same distribution of messages as $\np^*$ would see in a ``real''
interaction with $\nv$; thus, we have that the $(j,p)$ is an accepting
message (w.r.t., the challenge $r \oplus z_i$) with probability
$\epsilon$. Additionally, since $r \oplus z_i$
information-theoretically hides $i$ (as $r$ is completely random), we
have that the probability that $i = j$ is $\frac{1}{\ell_c}$ and
furthermore, the event that this happens is independent of whether the
message $(j,p)$ is accepting. We conclude that $\Adv^*$ convinces
$\Chal$ with probability $\frac{\epsilon(n)}{\ell_c}$, which concludes
the soundness proof.
\EPF

\subsection{\cPromise Distributional Problems}\label{sec:hardma}
We now conclude our main theorem that a hard-on-average language in
$\NP$ implies hard-on-average {\promise} distributional search
problem.

We first show that 2-round public-coin puzzles imply 2-round
(private-coin) puzzles with perfect completeness:
\BT
Suppose there exists $2$-round public-coin puzzle. Then there exists a
$2$-round private-coin puzzle with perfect completeness. 
\ET
\BPF
The theorem follows directly by applying our earlier proved results:
\BI
\item By Theorem~\ref{thm:2toperfect} (perfect completeness through
    adding a round), a 2-round public-coin puzzle implies a 3-round public-coin
    puzzle with perfect completeness.
    \item By Lemma~\ref{lem:collapse:var2} (round-collapse lemma), we
      conclude that either ioOWF exists, or there exists a
      2-round public-coin puzzle with perfect completeness.
      \item As ioOWFs trivially imply a 2-round (private-coin) puzzle,
        the theorem follows.
        \EI
        \EPF
 \noindent By observing that 2-round \emph{private-coin} puzzles
with perfect completeness are syntactically equivalent
to a hard-on-average \emph{\promise} distributional search problem,
and recalling that by Lemma~\ref{lem:hoa2puz}, aeHOA distributional
  $\NP$ problem implies a 2-round puzzle, we directly get the following corollary:
\BCR
 Suppose there exists a distributional $\NP$ problem $(L,\cD)$ that is
 aeHOA. Then, there exists a hard-on-average {\promise} distributional $\NP$ search problem.
 \ECR 
In other words, ``it isn't easier to prove efficiently-sampled statements that are
guaranteed to the true''.

\subsection{$\TFNP$ is Hard in Pessiland}
We next use the same approach to conclude that a hard-on-average
language in $\NP$ implies either (1) the existence of one-way
functions, or (2) the existence of a hard-on-average problem in
$\TFNP$.

\BT
 Suppose there exists a distributional $\NP$ problem $(L,\cD)$ that is
 aeHOA. Then, either of the following holds:
 \BI \item
 There exists a OWF;
 \item There exists some $\cR \in \TFNP$
   and some $\PPT$ $\cD$ such that $(\cR,\cD)$ is SearchHAO.
   \EI
\ET

\BPF
Again, the theorem follows by simply applying our earlier proved results:
\BI
\item From Lemma~\ref{lem:hoa2puz}, we have that an aeHOA distributional
  $\NP$ problem implies a 2-round almost-everywhere puzzle.
  \item By Theorem~\ref{thm:2toperfect} (perfect completeness through
    adding a round), this implies a 3-round almost-everywhere
    puzzle with perfect completeness.
    \item Applying
Lemma~\ref{lem:collapse:var2} (round-collapse, variant 2), we conclude that either
one-way functions exists, or there exists a 2-round public-coin puzzle with
perfect completeness.
\item Finally, by applying Lemma~\ref{lem:ioperfpuzTFNP}, a 2-round
  \emph{public-coin} puzzle with perfect completeness implies the existence of some $\cR \in \TFNP$
   and some $\PPT$ $\cD$ such that $(\cR,\cD)$ is SearchHAO.
\EI
\EPF
\noindent By replacing the use of Lemma~\ref{lem:collapse:var2} with
Lemma~\ref{lem:collapse:var1} (round-collapse, variant 1), we instead
get the following variants.
\BT
 Suppose there exists a distributional $\NP$ problem $(L,\cD)$ that is
 aeHOA. Then, either of the following holds:
 \BI \item
 There exists an ioOWF;
 \item There exists some $\cR \in \TFNP$
   and some $\PPT$ $\cD$ such that $(\cR,\cD)$ is aeSearchHAO.
   \EI
\ET

\section{Characterizing Non-trivial Public-coin Arguments}
\label{nontrivial.sec}
We finally apply our round-collapse theorem to arguments systems.

\paragraph{Non-trivial arguments}
We first define the notion of a non-trivial argument. Whereas such a
notion of a non-trivial argument has been discussed in the community
for at least 15 years, as far as we know, the first explicit
formalization in the literature appears in a recent work by Goldreich \cite{oded18}. We simply say
that an argument system is \emph{non-trivial} if it is not a proof
systems---i.e., the computation aspect of the soundness condition is
``real''. 
\BD[non-trivial arguments] An argument system $(P,V)$ for a language $L$ is called \emph{non-trivial} if $(P,V)$ is not an interactive proof system for $L$. 
\ED
We focus our attention on \emph{public-coin arguments}.
We show that the existence of any $O(1)$-round
public-coin non-trivial argument implies the existence of
distributional $\NPpoly$ problem that is nuHAO.

\BT\label{cor:ntarg}
Assume there exists a $O(1)$-round public-coin non-trivial argument for
some language $L$. Then, there exists a distributional $\NPpoly$ problem that is  nuHOA. 
\ET
\BPF
Consider some $k$-round non-trivial public-coin argument system
$(P,V)$. We show that this implies the existence of a $k$-round non-uniform puzzle. The theorem
next follows by applying Corollary~\ref{cor:kto2}. 
 
Since $(P,V)$ is not a proof system, there must exist a some
polynomial $p(\cdot)$, an unbounded prover $B$, and sequences $I = \{n_1,
  n_2, \ldots\}$ and $\{x_{n_i}\}_{i\in \NN}$ such that for all $i \in
  \NN$, $|x_{n_i}| = n_i, x_{n_i} \notin L$ yet $B$ convinces $V$ on
  common input $x_{n_i}$ with probability $\frac{1}{p(n_i)}$.

  Now consider the $k$-round non-uniform puzzle $\Chal$ that for each
  $n$ receives $(1,x_{n})$ as non-uniform advice if $n \in I$ and
  otherwise $(0,\bot)$. Given non-uniform advice $(b,x)$, $\Chal(1^n)$
  simply accepts if $b = 0$ and otherwise runs the verifier $V(x_n)$.
  We shall argue that $\Chal$ is a $(\frac{1}{p(n)}, \frac{1}{2p(n)})$
  puzzle which by Remark~\ref{rem:par} implies a puzzle. Completeness
  follows directly from the existence of $B$ (when $b = 0$, we have
  completeness 1 and otherwise, we have completeness $\frac{1}{p(n)}$
  by construction). To show soundness, notice that any non-uniform
  $\PPT$ adversary $\Adv^*$
  that breaks soundness of the puzzle with probability $\frac{1}{2p(n)}$ for all sufficiently large $n$,
  must in particular break it for infinitely many $n \in I$, and as
  such breaks the soundness of $(P,V)$ for infinitely many $x \in
  \{0,1\}^* - L$ with probability $\frac{1}{2p(|x|)}$, which contradicts
  the soundness of $(P,V)$.
\EPF
We next remark that the implication is almost tight. The existence of
a nuHOA problem in $\NP$ (as opposed to $\NPpoly$) implies a 2-round
non-trivial public-coin argument for $\NP$.

\BL
Suppose there exists a distributional $\NP$ problem $(L',\cD)$ that is
nuHOA . Then, for every language $L \in \NP$, there exists a
non-trivial 2-round public-coin argument for $L$ with an efficient prover.
\EL
\BPF
We first observe that by the same proof as for
Lemma \ref{lem:hoa2puz}, a nuHOA $\NP$ problem implies a 2-round
puzzle satisfying a ``weak'' completeness property, where completeness
only holds for infinitely many $n \in \NN$, but where soundness holds
also against non-uniform $\PPT$ algorithms. (Recall that in the proof
of Lemma \ref{lem:hoa2puz}, we only relied on the almost-everywhere
HOA property of the $\NP$ problem to ensure that completeness held for
\emph{all} sufficiently large input lengths.) We next simply combine
this ``weakly-complete'' puzzle with a standard $\NP$ proof for
$L$ to get a non-trivial 2-round argument for $L$. More precisely,
the verifier $V(x)$ samples the first message of the puzzle and sends
it to the prover; next the verifier accepts the prover's response if
it is either a witness for $x \in L$ (for some witness relation for
$L$), or if the response is a valid response to the
puzzle. The honest prover $P$ simply sends a valid witness for
$x$. Completeness of $(P,V)$ trivially holds. Soundness holds due to the
soundness of the puzzle (w.r.t. nu $\PPT)$. By the weak completeness
property of the puzzle, we additionally have that $(P,V)$ is not an
interactive proof (since there are infinitely many input lengths on
which an unbounded prover can find a puzzle solution and thus break soundness).
\EPF

We finally observe that the existence of $n^{\epsilon}$-round
non-trivial public-coin arguments is equivalent to $\PSPACE \not
\subseteq \Ppoly$. We remark that one direction (that non-trivial
arguments imply $\PSPACE \not
\subseteq \Ppoly$) was already previously proven by Goldreich
\cite{oded18}.

 \BT[informally stated]\label{thm:NTexist}
For every $\epsilon>0$, there exists an (efficient-prover) $n^\epsilon$-round non-trivial
public-coin argument (for $\NP$) if and only if $\PSPACE
\not\subseteq \Ppoly$.
\ET
\BPF
The ``only-if'' direction (which was already proven by Goldreich \cite{oded18}) follows just as the only-if direction
of Theorem~\ref{thm:pspace}.  The ``if'' direction follows by combining a
standard $\NP$ proof with the puzzle from Theorem \ref{thm:pspace} and requiring the prover to either provide the
$\NP$ witness, or to provide a solution to puzzle. 
\EPF

\paragraph{Round Collapse for Succinct Arguments}
We proceed to remark that the proof of our round-collapse theorem also
has consequences for succinct \cite{Kilian92} and universal
\cite{M00, BG02} argument systems.
\BT\label{lem:argcollapse}
Assume there exists a $k$-round public-coin (efficient-prover) argument system for $L$ with
communication complexity $\ell(\cdot)$, where $k$ is a constant. Then,
either non-uniform ioOWFs exists, or there exists a 2-round
public-coin (efficient-prover) argument for $L$ with communication complexity ${O}(\ell(n)\mathsf{polylog}(n))^{k(n)-1}$.
\ET 
\BPF
We apply the BM round-collapse transformation to the $k$-round
argument system $k-1$ times, and between each application repeat the
protocol in parallel $\log^2 n$ times (where $n$ is the length of the
statement to be proven). Completeness (also w.r.t. efficient provers) follows
directly from the classic proof of the BM round collapse
\cite{BM88}. To show soundness, as before, we consider a single
application of the round-collapse transformation.
Consider an adversary that breaks the
soundness of the $k-1$-round argument system with probability
$\epsilon(n)$ for infinitely many $\{x_n\}_n$; by
\cite{PassV12,HastadPWP10} such an adversary can be turned into an
adversary that break the soundness of a single of the $\log^2 n$
repetitions of the protocol obtained after the BM transformation with
probability $\eps'(n)>\frac{1}{2}$ for infinitely many $\{x_n\}_n$. If
non-uniform ioOWFs does not exist, then we can rely on the same
construction as in the proof of Lemma~\ref{lem:collapse} to construct
an adversary $B^*$ that breaks the $k$-round argument system for the
same statements $\{x_n\}_n$ with probability $\frac{1}{64}$, which
contradicts the soundness of the $k$-round argument.

Note that each step of the round-collapse transformation has a
multiplicative overhead of $O(\ell_a(n) \mathsf{polylog(n)})$  where
$\ell_a(n)$ bounds the length of the prover messages.
{Therefore, iterating the round collapse transformation $i$ times will result in a multiplicative overhead of $O((\ell(n) \mathsf{polylog(n)})^{i})$.}
\EPF
Theorem~\ref{lem:argcollapse} thus shows that the existence of a $O(1)$-round
succinct (i.e., with sublinear or polylogarithmic communication
complexity) public-coin argument systems can
either be collapsed into a 2-round public-coin succinct argument for
the same language (and while preserving communication complexity up to
polylogarithmic factors, as well as prover efficiency), or non-uniform
ioOWF exist.

It is worthwhile to also note that if the underlying $O(1)$-round protocol
satisfies some notion of \emph{resettable} \cite{CGGM00} privacy for the prover
(e.g., resettable witness indistinguishability (WI) or witness hiding (WH)
\cite{CGGM00,FS90}), then so will the resulting 2-round
  protocol.
(The reason we do not consider resettable
  zero-knowledge is that due to \cite{OW93} even just plain
  zero-knowledge protocols for non-trivial languages imply the
  existence of a non-uniform ioOWF; thus for resettable
  zero-knowledge, the result would hold vacuously assuming $NP \not
  \subseteq \BPP$. However, 
  it is not known whether (resettable) WI or
WH arguments for non-trivial languages imply non-uniform ioOWFs.)

\section{Acknowledgements}
We are grateful to Johan H\aa stad and Salil Vadhan for
discussions about non-trivial arguments back in 2005. We are also very
grateful to Eylon Yogev for helpful discussions.
\bibliography{crypto}
\bibliographystyle{alpha}
\appendix

\section{Some Theorems from Average-Case Complexity}\label{app:extendedavg}
In this section, we provide formal justifications for Lemmas~\ref{lem:privatetopublic}, \ref{lem:anydelta} and \ref{lem:searchtodecision}
We recall some previous results on average-case complexity relevant to our work. 
\BT[\cite{Trevisan05}]\label{thm:trevisan}
Suppose that there exists a $\NP$ language $L$ and polynomials
$\ell(\cdot)$
and $p(\cdot)$ such that $(L,\cU_\ell)$ is
$\frac{1}{p(n)}$-HOA (resp., $\frac{1}{p(n)}$-aeHOA and
$\frac{1}{p(n)}$-nuHOA). Then there exists a $\NP$ language $L'$ and
polynomial $\ell'(\cdot)$ such that $(L',\cU_{\ell'})$ is $\frac{1}{2}-\frac{1}{(\log n)^\alpha}$-HOA (resp., $\frac{1}{2}-\frac{1}{(\log n)^\alpha}$-aeHOA and $\frac{1}{2}-\frac{1}{(\log n)^\alpha}$-nuHOA).
The value $\alpha > 0$ is an absolute constant.
\ET



\BT[\cite{ImpagliazzoL90}]\label{thm:IL}
Suppose there exists a distributional $\NP$ search problem $(\cR,\cD)$ that is $\frac{1}{p(n)}$-SearchHOA (resp., $\frac{1}{p(n)}$-aeSearchHOA and $\frac{1}{p(n)}$-nuSearchHOA) for some polynomial $p(\cdot)$. Then there exists a search problem $\cR'$ and polynomials $\ell(\cdot)$ and $q(\cdot)$ such that $(\cR',\cU_\ell)$ is $\frac{1}{q(n)}$-SearchHOA (resp., $\frac{1}{q(n)}$-aeSearchHOA and $\frac{1}{q(n)}$-nuSearchHOA).
\ET

\BT[\cite{Ben-DavidCGL92}]\label{thm:BCGL}
Suppose that there exists a distributional $\NP$ search problem
$(\cR,\cU_\ell)$ that is $\frac{1}{p(n)}$-SearchHOA (resp.,
$\frac{1}{p(n)}$-aeSearchHOA and $\frac{1}{p(n)}$-nuSearchHOA) for some polynomials $p(\cdot)$ and
$\ell(\cdot)$.
Then there is a $\NP$-language $L'$ and polynomials $\ell'(\cdot)$ and $q(\cdot)$ such that $(L',\cU_{\ell'})$ is $\frac{1}{q(n)}$-HOA (resp., $\frac{1}{q(n)}$-aeHOA and $\frac{1}{q(n)}$-nuHOA). If we start with a distributional $\NP/\poly$ search problem $(\cR,\cU_\ell)$ that is $\frac{1}{p(n)}$-nuSearchHOA, then we obtain $L' \in \NP/\poly$ such that $(L',\cU_{\ell'})$ is $\frac{1}{q(n)}$-nuHOA.
\ET

Theorems~\ref{thm:trevisan},~\ref{thm:IL}~and~\ref{thm:BCGL} are stated in a slightly different form in \cite{Trevisan05,ImpagliazzoL90,Ben-DavidCGL92}. Namely, we highlight that the reduction is in fact ``length-regular''\footnote{A length-regular function $f: \{0,1\}^* \rightarrow \{0,1\}^*$ satisfies the properties that: (1) $|x| = |y| \Leftrightarrow |f(x)| = |f(y)|$,and (2) $|x| < |y| \Leftrightarrow |f(x)| < |f(y)|$ for any two strings $x$,$y$. We require the ``length-regular'' property for Turing (Cook) reductions where solving an instance $x$ on the target language requires queries the oracle only on instances of size $\ell(|x|)$ on the source language where $\ell$ is a non-decreasing function.} in that solving instances of size $\ell(n)$ in the target language helps solving instances of size $n$ in the source language. We will require this stronger property for the reductions to hold in the case of almost-everywhere hardness. 

\paragraph{Proof of Lemma~\ref{lem:anydelta}.} This follows immediately from Theorem~\ref{thm:trevisan}.

\paragraph{Proof of Lemma~\ref{lem:searchtodecision}.} Suppose there exists a distributional $\NP$-search problem
$(\cR,\cD)$ that is SearchHOA (resp., aeSearchHOA and nuSearchHOA). By
Theorem~\ref{thm:IL}, there exists a search problem $\cR'$ and
polynomials $\ell(\cdot), q(\cdot)$ such that $(\cR',\cU_\ell)$ is
$\frac{1}{q(n)}$-SearchHOA (resp., aeSearchHOA and nuSearchHOA). Next, by Theorem~\ref{thm:BCGL},  there
is a $\NP$-language $L'$ and polynomials $\ell'(\cdot)$ and
$q'(\cdot)$ such that $(L,\cU_\ell)$ is $\frac{1}{q'(n)}$-HOA
(resp. aeHOA and nuHAO), which
when combined with Theorem~\ref{thm:trevisan} yields a $\NP$ language
$L''$ and a polynomial $\ell''$ such that $(L',\cU_{\ell''})$ is
$\frac{1}{2}-\frac{1}{(\log n)^\alpha}$-HOA (resp., aeHOA and nuHAO). This implies that
$(L'',\cU_{\ell''})$ is HOA (resp., aeHOA and nuHOA).

\paragraph{Proof of Lemma~\ref{lem:privatetopublic}.} Suppose
$(L,\cD)$ is a distributional $\NP$ problem that is HOA (resp. a
distributional $\NPpoly$ problem that is nuHOA), then $(\cR,\cD)$ is
SearchHOA (resp., nuSearchHOA) where $\cR$ is the witness relation
corresponding to $L$. By Lemma~\ref{lem:searchtodecision}, we can
obtain a $\NP$ (resp. $\NPpoly$) language $L'$ and polynomial $\ell'$ such that $(L',\cU_{\ell'})$ is HOA (resp., nuHOA). 

\end{document}